\documentclass{article}
\usepackage{emulateapj}
\usepackage{onecolfloat}
\usepackage{graphicx}

\lefthead{Weiner, Sellwood \& Williams}
\righthead{Disk and Halo Mass of NGC 4123. II.}

\slugcomment{\it Accepted to ApJ}

\long\def\Ignore#1{\relax}

\newcommand\captsize{\scriptsize}

\newcommand\kms{\hbox{$\rm{km}~\rm{s}^{-1}$}}
\newcommand\ha{\hbox{H$\alpha$}}
\newcommand\hi{\hbox{H~I}}
\newcommand\mden{\hbox{${\rm M}_{\odot}~{\rm pc^{-3}}$}}
\newcommand\msurfden{\hbox{${\rm M}_{\odot}~{\rm pc^{-2}}$}}
\newcommand\kmskpc{\hbox{$\rm{km}~\rm{s}^{-1}~\rm{kpc}^{-1}$}}
\def\deg{\hbox{$^{\circ}$}}
\newcommand\msun{\hbox{${\rm M}_{\odot}$}}
\newcommand\lsun{\hbox{${\rm L}_{\odot}$}}

\newcommand\etal{{\it et al.}}
\newcommand\eg{{\it e.g.\ }}
\newcommand\ml{\hbox{$M/L$}}
\newcommand\mlratio{\hbox{$\Upsilon_b/\Upsilon_d$}}

\def\hii{H\,\,\textsc{ii}}
\def\lsun{\hbox{$L_{\odot}$}}

\def\mli{\hbox{$M/L_I$}}
\def\rlag{\hbox{$r_L$}}
\def\rlaga{\hbox{$r_L/a$}}
\def\omegap{\hbox{$\Omega_p$}}
\def\chisq{\hbox{$\chi^2$}}
\def\rchisq{\hbox{$\chi^2/N$}}

\def\beginfig{\begin{figure}}
\def\endfig{\end{figure}}
\def\beginfigtwo{\begin{figure*}}
\def\endfigtwo{\end{figure*}}

\def\begintab{\begin{table}}
\def\endtab{\end{table}}
\def\begintabtwo{\begin{table*}}
\def\endtabtwo{\end{table*}}

\begin{document}

\twocolumn[

\title{The Disk and Dark Halo Mass of the Barred Galaxy NGC 4123.\\
II. Fluid-Dynamical Models}

\author{Benjamin J.\ Weiner
}
\affil{Observatories of the Carnegie Institution of Washington,
813 Santa Barbara St, Pasadena, CA  91101}

\author{J.\ A.\ Sellwood and T.\ B.\ Williams}
\affil{Department of Physics and Astronomy, Rutgers University,
136 Frelinghuysen Rd., Piscataway, NJ 08854}


\begin{abstract}

We report a dynamical determination of the separate contributions of
disk and dark halo masses to the rotation curve of a spiral galaxy.
We use fluid-dynamical models of gas flow in the barred galaxy NGC
4123 to constrain the dynamical properties of the galaxy: disk \ml,
bar pattern speed, and the central density and scale radius of the
dark halo.  We derive a realistic barred potential directly from the
light distribution.  For each model we assume a value of the stellar
\ml\ and a bar pattern speed \omegap\ and add a dark halo to fit the
rotation curve.  We then compute the gas flow velocities with a 2-D
gas dynamical code, and compare the model flow patterns to a 2-D
velocity field derived from Fabry-Perot observations.  The strong
shocks and non-circular motions in the observed gas flow require a
high stellar \ml\ and a fast-rotating bar.  Models with $I$-band disk
\ml\ of 2.0 -- 2.5$h_{75}$, or 80 -- 100\% of the maximum disk value,
are highly favored.  The corotation radius of the bar must be $\leq
1.5$ times the bar semi-major axis.  These results contradict some
recent claimed ``universal'' galaxy disk/halo relations, since NGC
4123 is of modest size (rotation curve maximum 145~\kms, and $V_{\rm
flat}=130$ \kms) yet is quite disk-dominated.  The dark halo of NGC
4123 is less concentrated than favored by current models of dark halos
based on cosmological simulations.  Since some 30\% of bright disk
galaxies are strongly barred and have dust lanes indicating shock
morphology similar to that of NGC 4123, it is likely that they also
have high stellar \ml\ and low density halos. We suggest that luminous
matter dominates inside the optical radius $R_{25}$ of high surface
brightness disk galaxies.

\end{abstract}


 ]


\section{Introduction}

Extended rotation curves of spiral galaxies provide the 
most solid evidence for mass discrepancies on galactic
scales, requiring galaxies to be embedded in halos of
dark matter.  However, the relative contributions of
disk and halo to the mass of galaxies are not well known.
The rotation curve of an axisymmetric disk galaxy does not provide 
enough information to disentangle the disk and halo contributions to the 
mass distribution.
Even a well-sampled and spatially extended 
rotation curve can be fitted well by any 
combination of disk and halo from no disk to a ``maximum disk'' value (van 
Albada \etal\ 1985).  There is no consensus yet on whether maximum 
disks are preferred; we review arguments for and against in section 
\ref{sec-maxdisk}.
As a result, we do not know the relative importance of disk 
and halo, which hampers our understanding of major problems
in galaxy formation and dynamics, including the efficiency of
assembly of baryons into disks, the origin of the
Tully-Fisher relation, and the importance of instabilities
such as bars, spiral arms, and warps. 

Here we show that the disk-halo degeneracy can be broken in barred 
galaxies.  Two-dimensional velocity fields show that bars
drive non-circular streaming motions (see Weiner \etal\ 2000, Paper I).
The strength and location of the non-circular motions
are governed by the ellipticity of the potential,
hence the mass of the bar,
and, to a lesser extent, by the bar pattern speed, 
the angular rate of figure rotation of the bar.
In strongly barred galaxies, the bar is 
the dominant visible component in the inner galaxy, and 
the (unobservable) dark halo should be more rounded than the bar.  Thus 
the extra information in the non-circular motions can
determine the mass-to-light ratio (\ml) of the bar and, by 
extension, the mass of the stellar disk.

In this paper, we model
the non-circular flow pattern of gas in the disk of 
NGC 4123, which we obtained in Paper I from optical and radio emission line 
observations.  Since we observe only the line-of-sight velocity, we cannot 
reconstruct the full space velocity field of the gas, as we can for an 
axisymmetric galaxy.  Furthermore, shocks are observed in the gas
velocity field, so fluid dynamical models of the gas flow are
needed.  We compare these models against
the observed velocity field in order to derive the disk \ml.
The disk \ml\ determines the importance of the bar, hence the ellipticity
of the potential, and thus controls the strength of
non-circular motions and shocks: large shocks require a fairly
non-circular potential and hence a significant contribution from
the stellar bar.

We construct models for the galactic potential from our photometric 
observations and run fluid-dynamical simulations of the gas flow 
in a range of model potentials, for different disk \ml\ and bar pattern speeds 
\omegap.  We then compare the model velocity fields to that observed to find 
the most likely values of \ml\ and \omegap.
As in Paper I, we assume a 
distance to NGC 4123 of 22.4 Mpc, which is based on a Hubble constant of 
$75~\kms~\rm{Mpc}^{-1}$; all values of \ml\ in this paper are implicitly 
followed by a factor of $h_{75}$.  

\section{The disk-halo degeneracy and maximum disks}
\label{sec-maxdisk}

Rotation curves of axisymmetric disk galaxies which
extend beyond the optical disk clearly require dark matter
but do not contain enough information to distinguish between
disk-halo decompositions ranging from zero-mass to maximum
disks (van Albada \etal\ 1985; van Albada \& Sancisi 1986).
The ``maximum disk'' hypothesis, requiring that the disk be as massive as 
possible, was originally invoked simply for definiteness to
place a lower bound on the dark mass.  
We review here some arguments for and against maximal disks
(see also the reviews of Bosma 1999 and Sellwood 1999).


Non-axisymmetric structure provides some information about disk \ml.
Two-armed spiral patterns are common in disks,
and excessive halo mass suppresses two-armed spiral structure,
favoring disks 50\% -- 100\% of maximum mass
(Athanassoula, Bosma \& Papaioannou 1987).
The maximum disk or even ``no halo'' assumptions
can reproduce the overall shape of the rotation curve inside the 
optical disk (\eg\ Kalnajs 1983; van Albada \etal\ 1985; Kent 1986; 
Begeman 1987; Freeman 1992; Palunas 1996; Broeils \& Courteau 1997).
Small-scale ``bumps and wiggles'' in the rotation curve 
do not offer compelling support for maximum disks (van der Kruit 
1995; Palunas 1996).  
Quillen \& Frogel (1997) integrated periodic orbits in
potentials for the ringed barred galaxy NGC 6782 and concluded that its 
disk should have an \ml\ of $75 \pm 15 \%$ of maximum
(see also Quillen, Frogel \& Gonzalez 1994 for application to a bar).

Dynamical friction from a heavy halo will slow down a bar
(Weinberg 1985; Debattista \& Sellwood 1998);
bars are slowed down rapidly in even mildly submaximal disks.
There are several reasons to believe that bars
rotate quickly (see the review of Elmegreen 1996), and two bars have
been observed to be fast rotators (Merrifield \& Kuijken 1995; 
Gerssen, Merrifield \& Kuijken 1999).
The bar of the Milky Way is also best modeled as
a fast rotator (Weiner \& Sellwood 1999; Fux 1999).

Bottema (1993, 1997) has argued that stellar velocity dispersions imply 
that disks supply only about 63\% of the maximum rotation velocity, 
so that disk \ml\ is 40\% of maximum (see also Fuchs 1999).  
However, Bottema's analysis rests on a chain of assumptions 
and the upper bound can be relaxed (Bosma 1999).
While the Oort limit from the vertical stellar 
velocity dispersion has a long history as a measurement of disk mass 
(\eg\ Oort 1932; Kuijken \& Gilmore 1991)
it is still unclear whether even the Milky Way disk is maximal or not
(Sackett 1997; Binney \& Merrifield 1998, chapter 10).
Its application to
external galaxies has two major complications: (1) we cannot 
measure both vertical scale height and vertical velocity 
dispersion in any individual galaxy; 
(2) integrated-light measurements of the vertical velocity 
dispersion will be dominated by young stars, which have a 
lower dispersion in the Milky Way (see Binney \& Merrifield 1998, 
chapter 10; but cf.\ Fuchs 1999).

Maller \etal\ (2000) used gravitational lensing by an edge-on
disk galaxy at $z=0.41$ to measure its potential, concluding that
a maximal disk is ruled out, though a model with a massive
bulge and sub-dominant halo is allowed.  The lack of kinematic
information makes attempts at rotation curve decomposition
preliminary.

Courteau \& Rix (1999) show that if galaxies have maximal disks, 
there should be a relationship between scale length and 
the peak of the disk rotation curve $V_{2.2}$, 
predicting a correlation between scale length 
and residual from the Tully-Fisher relation $\Delta V_{2.2}$: smaller disks 
should have a higher rotation width at a given luminosity.  They find no 
such correlation between the residuals from the velocity width--magnitude 
and scale length--magnitude relations, implying the disk contribution
to rotation width is small.

In searching for a correlation between scale length and rotation width 
at a fixed luminosity, there is a surface brightness dependence: 
larger disks will have lower surface brightness.
As Courteau \& Rix note, the lack of correlation
demonstrates that surface brightness is not a second parameter in the TF 
relation.  LSB disks do indeed lie on the same TF relation 
as HSB disks (Zwaan \etal\ 1995; Sprayberry \etal\ 1995), although
a ``baryonic TF'' relation may be required
(McGaugh \etal\ 2000; O'Neil, Bothun \& Schombert 2000).
Lower surface brightness disks appear to be quite halo-dominated
(\eg\ de Blok \& McGaugh 1996, 1997; Swaters, Madore \& Trewhella
2000), and should, in fact, violate the scale length--TF residual
relation expected for maximal disks.
However, to preserve the TF relation, either disks are
negligible in all galaxies or the relation between halo/disk ratio 
and surface brightness must be fine-tuned.

The lack of correlation between
scale length and $\Delta V_{2.2}$ indicates that there is no
feature in the rotation curve to distinguish putatively disk-dominated
and halo-dominated portions of the rotation curve.
This fine-tuning
is related to the well-known ``disk-halo conspiracy'' noted by 
Bahcall \& Casertano (1985): if disks are maximal, then disk and halo
rotation curves must be of similar amplitude to maintain
flat rotation curves (but see Casertano \& van Gorkom 1991).
The mystery in the end is why LSB and HSB disks lie on the same 
Tully-Fisher relation.

The best way to resolve the disk-halo degeneracy and its attendant
problems, and to understand the origin of the Tully-Fisher relation,
is to actually measure the degree of 
maximality of disks of galaxies over a range of velocity width and surface 
brightness.  We are now attempting to make this measurement through 
observations of a number of barred galaxies.  In this paper,
we carry out this procedure for NGC 4123.

\section{Modeling I. The stellar mass distribution}

We wish to estimate the gravitational potential of the
three-dimensional stellar mass distribution of the galaxy given a
two-dimensional surface brightness distribution.  There are three
potential difficulties: spatial variations in \ml\ or extinction,
deprojection to estimate the 3-D luminosity density from its
projection into the 2-D plane of the sky, and conversion of luminosity
density into mass density.

In fact, we do not need the full 3-D gravitational potential, but only its 
derivatives in the galaxy mid-plane -- vertical gradients can be 
neglected.  This is because the emission-line kinematic tracers we observe 
arise in the gas phase, which lies in a thin layer.  
(The \hi\ disk can be warped and/or flaring at large 
radius, but these phenomena generally occur outside the optically luminous
disk of the galaxy -- see the review by Binney (1991).)  This requirement 
considerably simplifies the task of deprojection.

As argued in Paper I, color and extinction variations are mild in 
our $I$ band image, and correcting for them would be quite 
model-dependent; we therefore neglect them here.

Projection is of course formally degenerate, and impossible to invert 
uniquely.  One approach is to fit a 3-D parametric model to the data by 
projecting it into 2-D.  Prolate Ferrers ellipsoids are the most 
convenient analytical forms with easily derived bar-like potentials, and 
several previous gas-dynamical studies of barred galaxies have used them 
to model bars.  Duval \& Athanassoula (1983) used a $n=0$ (uniform 
density) Ferrers bar in their study of NGC 5383, Regan \etal\ (1997) 
used a $n=1$ Ferrers bar to model NGC 1530, and we have
used a $n=1$ Ferrers bar to model the Milky Way (Weiner \& Sellwood 1999).
Lindblad \etal\ (1996), on the other hand,
used a photometric model less dependent on an 
analytical form to derive a potential for their gas-dynamical simulations, 
rectifying their $J$-band image of NGC 1365 to face-on and
decomposing it into the first several even Fourier components.

A disadvantage of the analytical approach is evident from the $I$-band 
image of NGC 4123 presented in Paper I.  The bar in this galaxy has a 
complex shape which does not lend itself to modeling by simple analytic 
expressions.  It has two components, both of which are clearly 
non-ellipsoidal.  The high surface brightness central component is 
distinctly rectangular, while the elongated component is also quite boxy 
and possibly broadens, if anything, towards the ends, where an ellipsoid 
would taper.  We are therefore motivated to construct a model which is 
independent of analytic expressions for the luminosity density.
Our approach is inspired by Quillen, Frogel \& Gonzalez
(1994), who estimated the 
potential of a face-on barred galaxy in a model-independent way by 
integrating over the actual light distribution.  We wish to apply the same 
method to an inclined galaxy, which we must first deproject to 
face on.  

\subsection{Deprojection}

Disk galaxies are relatively thin (\eg\ van der Kruit \& Searle 1982) and a 
simple stretch of our 2-D image should give us a reasonable approximation to 
a face-on view of the galaxy.  As the galaxy has a finite scale height, 
which is probably not the same for all components, such a deprojection will 
introduce biases which we address in this section.

There are a number of foreground stars which must be removed before the 
rectification process.  We masked out circular apertures around each star. 
NGC 4123 is very close to bisymmetric, especially in the bar region, so we 
replaced the missing data in the masked regions with the data from the 
corresponding regions after a 180\deg\ rotation.  Since the galaxy is so 
bisymmetric, we chose to conserve computing resources by running our 
simulations on a half-plane and enforcing bisymmetry (see Section 
\ref{sec-fluidmod}).  We constructed a strictly bisymmetric image by 
rotating the image 180\deg\ and averaging the rotated and original images.
We assume the 
inclination $i=45\deg$, derived from our kinematic data in Paper I,
and simply stretch the galaxy image along the projected minor 
axis by the factor $1/{\rm cos}~i$ to produce a face-on image of 
surface brightness, while conserving total luminosity.

We selected NGC 4123 for this study in part because it does not have 
a large spheroidal bulge.  In fact, the only spheroidal component present 
is a unresolved point-like source at the very center of the galaxy (Paper 
I), having a luminosity $1.8 \times 10^8 \lsun$ in the 
$I$-band. We fit this source independently and remove it before deprojection.

The finite thicknesses of the disk and bar will introduce inaccuracies
into our deprojection.  The typical scale height of a disk is a few 
hundred parsecs, but the vertical extent of the bar may be up to
2--3 times greater.
Bars in simulations develop pronounced ``peanut'' shapes 
(Combes \& Sanders 1981; Raha \etal\ 1991) which have larger scale heights 
than the disk, and the box/peanut bulges seen in 
some edge-on galaxies are associated with bars (Kuijken \& Merrifield 
1995; Bureau \& Freeman 1999).  
It is possible that the peanuts and boxes occur at the bar center, but that 
the bar is thinner further out along its length.  Dettmar \& Barteldrees 
(1990) showed that some edge-on galaxies with box and peanut bulges have 
an additional thin component which they suggest is a bar.  
NGC 4123 could be such an object, given the 
two-component structure of its bar (Paper I).

The bias introduced by rectification is significant only when a structure 
has a scale length $s$ along the deprojected minor axis of the galaxy that 
is comparable to its vertical scale height $z_0$.  At an inclination of 
$i=45\deg$, for example, erroneously rectifying a sphere with $s=z_0$ 
would stretch it along the galaxy minor axis by a factor of 1.414, a 
41\% error in its linear extent.  However, a spheroid with a modest 2:1 
flattening ($s=2z_0$), once rectified, would be erroneously stretched by a 
factor of just 1.12, and for an object with 3:1 flattening the stretch
is too large by just 5\%.  These factors are only marginally different for 
non-ellipsoidal cross-sections.

As noted above, the boxy structure of NGC 4123 inside $\sim 15\arcsec$ 
radius could be thickened into a box- or peanut-type bulge.
Our models may not be reliable in the 
very center for this reason, although the bulge is
unlikely to be as thick as 1:1.  
Outside the central boxy structure, the bar is 
probably about as thin as the disk, and the rectification bias should be 
minimal.   The elongated part of the bar has a projected scale length 
of $s_{\rm proj} \sim 16\arcsec$ on the plane of the sky, parallel to the 
galaxy minor axis (not the bar minor axis).
The deprojected length $s$ in the plane of the galaxy, parallel to the 
galaxy minor axis is greater, $s \simeq 1.4 s_{\rm proj} = 2.4$ kpc. 
If the bar has a scale height typical of disk galaxies, $z_0 \sim 400 $~pc, 
as we have assumed, then the erroneous stretching induced
by rectification is very small.  Rectification stretches along
the galaxy minor axis, not the bar minor axis, which works in our favor.

\subsection{Calculating the potential}

In order to calculate the gravitational field from this face-on image, we 
must make two additional assumptions: (1) some relation between light and 
mass and (2) a form for the vertical structure of the disk.  

We follow usual practice by assuming that light is directly proportional 
to mass -- that is, a constant \ml\ throughout the galaxy.  This simplest 
possible assumption is generally reasonable in the inner parts 
of galaxies (see Kent 1986), as evidenced by weak color gradients
and the success of maximum disk 
models with constant \ml.  Barred galaxies generally have shallower 
abundance gradients than unbarred galaxies (\eg\ Martin \& Roy 1994;
Zaritsky, Kennicutt \& Huchra
1994), which suggests they are well mixed and any 
gradient in \ml\ must also be fairly shallow.

We considered the effect of a radially varying \ml\ in Paper I.  Even if the 
outer disk \ml\ at 15 kpc is just half that of the bar, the effect on the 
disk contribution to the rotation curve is just 5\% at 15 kpc. Our modeling 
procedure is sensitive to the \ml\ inside the bar, and varying the disk \ml\ 
outside the bar makes little difference to the derived disk and halo masses.

The potential and accelerations in the midplane of a finite-thickness disk 
are weaker than those from a razor-thin disk.  We assume a vertical 
distribution of the common form $\rho(z) \propto {\rm sech}^2(z/2z_0)$, and 
a scale height of $z_0 = 200$~pc -- similar to that found in 
edge-on disk galaxies (van der Kruit \& 
Searle 1982; de Grijs \& van der Kruit 1996; de Grijs 1997)
and slightly smaller than in the Milky Way, since NGC 4123 is smaller
(Mihalas \& Binney 1981).
The scale height is effectively a smoothing length for the potential.
With the $z$-distribution given, 
we calculate the potential and the accelerations at 
every point in the midplane, using a Fourier transform method 
to convolve a Green's function with the surface density distribution
(Hockney 1965).

The rectified $I$-band image from our CTIO 0.9-meter photometry (Paper I) 
taken in 1.2\arcsec\ seeing has 0.39\arcsec\ pixels; in order to keep 
the cell size and number of cells in the simulation grid reasonable, we 
binned it $2\times2$ to make the pixel size 0.78\arcsec, or 84.7~pc.  We 
then used the FFT algorithm to generate the $x$- and $y$-acceleration 
components on a $1024 \times 1024$ (87 kpc square) grid, although we 
use only a $256 \times 512$ subsection of this grid for our 
simulations; the large grid ensures that forces from mass outside
the subsection are calculated accurately.  Since the FFT grid is larger 
than the field of the CTIO observations, we extrapolated the disk surface 
brightness profile.  The extrapolation contains just 2\% of the 
total disk luminosity and is consistent with the observed surface brightness 
profile from our wider-field Las Campanas photometry.

We tested the effect of changing the assumed
scale height $z_0$ by factors of up to 2.
The effects on the velocity jump across the bar shock
(see Figure~\ref{fig-cutml}) 
are fairly small; lowering the scale height makes little difference,
while increasing the scale height to 800 pc weakens the predicted 
velocity jump slightly, which would require a slightly higher bar 
\ml\ to match the observations (see Section \ref{subsec-slit}).  
The assumed scale height has little effect on our conclusions
based on the strength of the velocity jump (since we conclude
that the bar \ml\ is high).
However, the smoothing due to the scale height does affect
the models at $R < 1$ kpc; the streaming
motions at the inner Lindblad resonance (ILR) weaken with increasing 
scale height.  Therefore the models are not robust inside the ILR
(see Section~\ref{sec-comparison}).

The method also assumes that $z_0$ is constant over
the bar and disk. As argued above, 
only the inner 10\arcsec--15\arcsec\ of the bar can be substantially 
thicker than the disk, and the effect is small
outside the innermost few grid cells; but 
inside $\sim 10$ grid cells, the approximate size of the ILR,
the potential could be affected by the uncertainty in scale height.

\section{Modeling II. Additional mass components}
\label{sec-potential}

The total gravitational field must  account for components of the mass 
distribution other than the stellar disk; these include the nucleus, 
gas disk, and dark halo of NGC 4123.


We removed the central point-like source of 
$1.8 \times 10^8 \lsun$ from the image before the 
rectification described above.  Its \ml\ is uncertain, 
since it is blue and an emission-line source,
and is likely to have a young stellar 
population or even nonstellar luminosity, both suggesting a 
low \ml.  Conversely, there could be a black hole at the center of NGC 
4123, although upper limits on the black hole mass to bulge mass relation
make it highly unlikely that a black hole in such a 
modest-sized galaxy could have a mass as high as $10^8 \msun$ 
(Kormendy \& Richstone 1995).

We modeled the nucleus/bulge as a spherical Gaussian distribution, $\rho(r) 
\propto \exp(-r^2/{r_c}^2)$, with a scale radius of 200 pc.  We chose this 
scale radius to be softer than implied by the optical observations
to keep the forces from varying too strongly 
over the innermost few grid cells, which would cause large numerical 
diffusion effects.  The Gaussian profile drops off rapidly, so
the softening affects the forces in only the innermost cells.

Since the \ml\ of the nucleus is uncertain, we ran sets of models with three 
different prescriptions for the \ml, setting \mlratio, the ratio of 
nucleus/bulge to disk \ml, equal to either 0, 0.5, or 1.0. The nuclear mass 
influences the gas flow by changing the strength or size of the inner 
Lindblad resonance 
(see Athanassoula 1992; Weiner \& Sellwood 1999).  While the different 
choices of nuclear mass do affect the flow in the inner few grid cells,
the dynamical parameters disk \ml\ and bar pattern speed \omegap\ are 
robust against variations in nuclear mass, as discussed further in 
Section~\ref{sec-comparison}.

Our 21 cm observations of NGC 4123 (Paper I) revealed the presence of a 
large, extended \hi\ disk, with a hole of lower surface density in its 
center.  The total mass of atomic gas, including helium, is $1.06 
\times 10^{10} \msun$.  In Figure 8 of Paper I, we show the small 
contribution of the azimuthally averaged atomic gas to the rotation curve. 
In this paper, we add this axisymmetric contribution to that of the 
now non-axisymmetric stellar disk.   We do not have a map of the
molecular gas distribution; as discussed in Paper I, single-dish 
observations show that the
contribution from molecular gas to the rotation curve
is quite small, and we neglect it.

In Paper I, we found that the observed 1-D rotation curve of NGC 4123 can 
be fitted satisfactorily for a wide range of disk \ml\ by adding quite 
simple DM halo models.  This property is true for many disk galaxies with 
extended 21 cm rotation curves and lies at the heart of the disk-halo 
degeneracy.

The DM halos used in the 2-D models of the velocity field considered here 
are identical to those determined by the fit to the 21 cm rotation curve 
for the corresponding axisymmetric \ml\ models in Paper I.  As there, we 
study two sets of models having two different popular halo density 
profiles:
\begin{equation}
\rho_h(r) = \cases{ \displaystyle \rho_{0} \frac{1}{1 + (r/r_c)^2}, & 
pseudo-isothermal; \cr
\displaystyle \rho_{s} \frac{4{r_s}^3}{r(r+r_s)^2}. & NFW-type. \cr}
\end{equation}
The density profile of the first set has the pseudo-isothermal form
while that of the second set is an ``NFW-type'' power-law profile (Navarro, 
Frenk, \& White 1996) discussed more fully in Paper I.

For a given disk \ml, the best-fit pseudo-isothermal and NFW-type halo 
rotation curves both fit the axisymmetric rotation curve well.  
Since they are fitting the same data, the two halo contributions 
to the rotation curve are quite similar and 
substituting one halo profile for the other makes 
little difference to the gas flow pattern inside the bar region.  In fact, 
the preferred values of the main parameters, disk \ml\ and \omegap, 
scarcely differ for the two possible halos (see  
Section~\ref{sec-comparison}).

We have assumed spherical halos.  As noted in Paper I, a flattened halo 
with a slightly different density distribution could produce essentially 
the same rotation curve.  Since our data are restricted to the midplane of 
the galaxy, they do not constrain the halo flattening, nor does the halo 
flattening affect the results.  If the halo were non-axisymmetric, it 
could conceivably contribute to the observed non-circular streaming 
motions.  However, to have a significant effect on
our results, the halo would have to 
be both quite prolate inside the bar radius, where the non-circular motions 
are large, and axisymmetric further out so as to not disturb the 
apparently circular flow pattern in the outer disk -- at which point it 
would be part of the bar, for all practical purposes.

\section{Modeling III. Fluid dynamical models}
\label{sec-fluidmod}

\subsection{The model parameters}

The disk \ml\ is the main parameter for the gravitational potential, and 
the nuclear mass and dark halo profile are subsidiary parameters. To run a 
simulation of the gas dynamics, we also need to know the bar pattern speed 
\omegap.  For a given potential, the value of \omegap\ also determines the 
radius \rlag\ of the Lagrange point L$_1$ of the bar; a faster-rotating 
bar corresponds to a smaller Lagrange radius.  A particle at the L$_1$ 
point would rotate at the same angular speed as the bar, and so \rlag\ is 
the corotation radius.  Two models with the same \omegap\ and different 
\ml\ have similar, but not exactly equal, values of \rlag, 
because the value of \rlag\ depends somewhat on the shape of the 
potential.

We parametrized our models by \ml\ and \omegap, and calculated \rlaga, the 
ratio of the Lagrange radius to the bar semi-major axis, for each model.  
The ratio \rlaga\ is commonly used to characterize the bar pattern speed 
because it generalizes across galaxies of different rotation curve height and 
bar size.  Bars with $\rlaga \la 1.5$ are generally considered to be 
``fast-rotating.'' The elongated stellar orbits which are generally 
presumed to make up the bar are confined inside the well of the effective 
potential which extends to L$_1$, so $\rlaga > 1.0$ is favored (\eg\ 
Teuben \& Sanders 1985; Binney \& Tremaine 1987). See Elmegreen (1996) for 
a review of the issues surrounding pattern speeds and their determination 
in barred galaxies.

A ``set'' of models includes a full range of disk \ml\ and bar pattern 
speed \omegap.  Each set is 88 models with all combinations of 8 values of 
disk \ml\ from \{1.0, 1.5, 1.75, 2.0, 2.25, 2.5, 2.75, 3.0\}, and 11 
values of \omegap\ from \{10, 12, 14, 16, 18, 19, 20, 21, 22, 23, 24\} 
\kmskpc.

In order to check the dependence of the models on the subsidiary 
parameters \mlratio\ and dark halo profile, we ran six sets of models 
covering the three choices of \mlratio\ \{0.0, 0.5, 1.0\} and the two halo 
profiles, isothermal and NFW-type.  We do not expect these subsidiary 
parameters to affect the models' fit to the data nearly as strongly as the 
main parameters, \ml\ and \omegap.  Conversely we do not expect to make a 
robust determination of \mlratio\ and halo profile from the models.  
In total, 528 models were run.

The models of the set with isothermal halos and $\mlratio=0.5$, which 
contains the best-fitting model, are listed by \ml\ and \omegap\ in Table 
\ref{table-params}.

The \ml\ range covers from a $\ml = 1.0$ disk, which is dominated by the 
dark halo at nearly all radii, to a super-maximal disk at the high end of 
$\ml = 3.0$ (see Figure 8 of Paper I).  The \omegap\ range covers from 
quite slow-rotating bars to super-fast bars with $\rlaga < 1$.


\begintabtwo[ht]
\captsize
\begin{center}
\hbox{
\hspace{0.75truein}
\begin{tabular}{ccrrr}
\tableline\tableline


 & & & \rchisq\phm{00} & \rchisq\phm{00}  \\
 $M/L_I$ & \omegap & \rlaga & 
     raw errors & + 8~\kms \\
 & & & & dispersion \\
\tableline

1.00 & 10.0 & 2.38 & 102.30 &  29.83  \\
1.00 & 12.0 & 2.04 &  55.21 &  14.74  \\
1.00 & 14.0 & 1.76 &  14.22 &   4.13  \\
1.00 & 16.0 & 1.53 &  12.79 &   3.73  \\
1.00 & 18.0 & 1.34 &  14.76 &   4.75  \\
1.00 & 19.0 & 1.25 &  16.00 &   5.15  \\
1.00 & 20.0 & 1.16 &  17.57 &   5.56  \\
1.00 & 21.0 & 1.07 &  19.44 &   6.03  \\
1.00 & 22.0 & 0.92 &  21.36 &   6.52  \\
1.00 & 23.0 & 0.85 &  22.97 &   6.92  \\
1.00 & 24.0 & 0.80 &  24.03 &   7.16  \\
1.50 & 10.0 & 2.48 &  84.42 &  24.14  \\
1.50 & 12.0 & 2.14 &  41.26 &  11.30  \\
1.50 & 14.0 & 1.86 &  21.04 &   6.01  \\
1.50 & 16.0 & 1.63 &  10.89 &   3.56  \\
1.50 & 18.0 & 1.47 &  10.64 &   3.67  \\
1.50 & 19.0 & 1.36 &  11.73 &   3.97  \\
1.50 & 20.0 & 1.28 &  13.36 &   4.38  \\
1.50 & 21.0 & 1.20 &  15.17 &   4.83  \\
1.50 & 22.0 & 1.11 &  16.56 &   5.20  \\
1.50 & 23.0 & 0.88 &  17.21 &   5.34  \\
1.50 & 24.0 & 0.83 &  17.51 &   5.37  \\
1.75 & 10.0 & 2.51 &  68.50 &  19.21  \\
1.75 & 12.0 & 2.16 &  36.06 &  10.25  \\
1.75 & 14.0 & 1.89 &  10.24 &   3.09  \\
1.75 & 16.0 & 1.67 &   7.06 &   2.50  \\
1.75 & 18.0 & 1.49 &   8.44 &   2.99  \\
1.75 & 19.0 & 1.40 &   9.70 &   3.34  \\
1.75 & 20.0 & 1.31 &  10.74 &   3.62  \\
1.75 & 21.0 & 1.23 &  10.97 &   3.69  \\
1.75 & 22.0 & 1.14 &  10.88 &   3.64  \\
1.75 & 23.0 & 0.85 &  11.11 &   3.64  \\
1.75 & 24.0 & 0.78 &  11.34 &   3.66  \\
2.00 & 10.0 & 2.52 &  74.36 &  20.18  \\
2.00 & 12.0 & 2.18 &  22.99 &   6.81  \\
2.00 & 14.0 & 1.91 &   7.09 &   2.40  \\
2.00 & 16.0 & 1.69 &   6.52 &   2.39  \\
2.00 & 18.0 & 1.51 &   5.89 &   2.11  \\
2.00 & 19.0 & 1.43 &   5.63 &   1.98  \\
2.00 & 20.0 & 1.33 &   5.23 &   1.87  \\
2.00 & 21.0 & 1.24 &   5.27 &   1.86  \\
2.00 & 22.0 & 0.86 &   5.65 &   1.97  \\
2.00 & 23.0 & 0.79 &   5.92 &   2.06  \\
2.00 & 24.0 & 0.71 &   6.33 &   2.14  \\
\tableline
\end{tabular}

\hspace{0.3truein}
\begin{tabular}{ccrrr}
\tableline\tableline

 & & & \rchisq\phm{00} & \rchisq\phm{00}  \\
 $M/L_I$ & \omegap & \rlaga & 
     raw errors & + 8~\kms \\
 & & & & dispersion \\
\tableline
2.25 & 10.0 & 2.53 &  70.29 &  19.56  \\
2.25 & 12.0 & 2.18 &  22.89 &   7.54  \\
2.25 & 14.0 & 1.93 &  12.56 &   4.67  \\
2.25 & 16.0 & 1.70 &   7.41 &   2.64  \\
2.25 & 18.0 & 1.53 &   6.27 &   2.23  \\
2.25 & 19.0 & 1.46 &   4.81 &   1.84  \\
2.25 & 20.0 & 1.35 &   3.54 &   1.40  \\
2.25 & 21.0 & 1.26 &   3.78 &   1.42  \\
2.25 & 22.0 & 0.85 &   4.22 &   1.55  \\
2.25 & 23.0 & 0.77 &   4.59 &   1.64  \\
2.25 & 24.0 & 0.68 &   4.87 &   1.67  \\
2.50 & 10.0 & 2.54 &  86.60 &  26.20  \\
2.50 & 12.0 & 2.20 &  89.74 &  25.09  \\
2.50 & 14.0 & 1.96 &  40.68 &  12.66  \\
2.50 & 16.0 & 1.73 &  20.93 &   7.16  \\
2.50 & 18.0 & 1.55 &   6.53 &   2.32  \\
2.50 & 19.0 & 1.48 &   8.50 &   3.03  \\
2.50 & 20.0 & 1.39 &   5.36 &   2.10  \\
2.50 & 21.0 & 1.31 &   5.11 &   1.87  \\
2.50 & 22.0 & 0.87 &   5.07 &   1.82  \\
2.50 & 23.0 & 0.80 &   4.33 &   1.60  \\
2.50 & 24.0 & 0.71 &   4.36 &   1.55  \\
2.75 & 10.0 & 2.54 & 129.66 &  36.64  \\
2.75 & 12.0 & 2.22 & 138.08 &  38.19  \\
2.75 & 14.0 & 1.99 &  62.56 &  19.22  \\
2.75 & 16.0 & 1.77 &  36.94 &  11.44  \\
2.75 & 18.0 & 1.58 &  17.23 &   6.12  \\
2.75 & 19.0 & 1.51 &   7.88 &   2.86  \\
2.75 & 20.0 & 1.46 &   9.83 &   3.22  \\
2.75 & 21.0 & 1.36 &   7.68 &   2.82  \\
2.75 & 22.0 & 1.28 &   6.93 &   2.48  \\
2.75 & 23.0 & 0.85 &   6.13 &   2.21  \\
2.75 & 24.0 & 0.78 &   5.46 &   1.93  \\
3.00 & 10.0 & 2.57 & 196.67 &  52.13  \\
3.00 & 12.0 & 2.25 & 186.90 &  51.66  \\
3.00 & 14.0 & 2.04 &  80.47 &  23.89  \\
3.00 & 16.0 & 1.84 &  48.29 &  14.96  \\
3.00 & 18.0 & 1.63 &  37.61 &  11.50  \\
3.00 & 19.0 & 1.55 &  22.39 &   7.45  \\
3.00 & 20.0 & 1.49 &  14.74 &   5.48  \\
3.00 & 21.0 & 1.43 &  10.49 &   3.39  \\
3.00 & 22.0 & 1.34 &  11.43 &   3.97  \\
3.00 & 23.0 & 1.27 &  10.22 &   3.31  \\
3.00 & 24.0 & 0.85 &   7.84 &   2.75  \\
\tableline
\end{tabular}
}
\end{center}

\caption{Simulation parameters and goodness of fit}
\label{table-params}
\tablecomments{
\captsize
Simulations for the set of models with isothermal halos and 
$\mlratio=0.5$, listed by their parameters \mli\ in Column 1 and \omegap\ 
in Column 2.  Column 3 is the value of \rlaga\ implied by \omegap.
Column 4 is the reduced \chisq\ for the models' fit to the velocity 
data using purely statistical errors (see Section \ref{sec-comparison}) 
and Column 5 is the reduced \chisq\ when an additional 8 \kms\ 
dispersion is added to the error budget.
}
\endtabtwo


\subsection{The fluid dynamical code}

We used a two-dimensional grid-based gas dynamical code (kindly provided 
to us by E. Athanassoula) to simulate the gas flow in the models of NGC 
4123.  The FS2 code is a second-order, flux-splitting, Eulerian, grid code 
for an isothermal gas in an imposed gravitational potential, originally 
written by G. D. van Albada to model gas flow in barred galaxy potentials 
(van Albada 1985).  Athanassoula (1992) used it to study gas flow patterns 
in a variety of model galaxy potentials.  We have modified it 
to suit our approach, but the heart of the code remains 
the finite-difference scheme described by van Albada (1985).

By its nature, the code approximates the interstellar medium as an Eulerian 
fluid, smooth on scales of the grid cell size.  Without some idealization it 
is hopeless to simulate the extremely complex dynamics of the multiphase 
ISM.  Various methods including grid codes, particle hydrodynamics,
and sticky-particle codes have been advocated (see Weiner \& Sellwood 1999).
Essentially, applying the Euler equations to the ISM simply 
asserts that the ISM has a pressure or sound speed defined in a 
coarse-grained sense, over scales greater than the code's resolution.

The smoothed particle hydrodynamics (SPH) 
models of Englmaier \& Gerhard (1997), who ran a simulation under the same 
conditions as used by Athanassoula (1992) with the FS2 Eulerian grid code, 
yielded very similar results. 
There are some small differences between the results of the grid
and SPH codes, chiefly due to the different tradeoffs in
resolution in Eulerian and Lagrangian schemes.  However, the
chief features, such as the strength of the shocks in the bar
(see Section \ref{sec-comparison}), are quite similar.
This comparison reassures us that the simulation results are not 
dependent on the particular fluid-dynamical algorithm.  

The code does not include the self-gravity of the gas, although the 
axisymmetrized rotation curve due to the atomic gas is included as part of 
our model for the potential, discussed above in Section 
\ref{sec-potential}.  The gas self-gravity is negligible for this study, 
since the gas surface density is considerably lower than that of the 
stellar disk and of the halo.   This is especially so in the bar region of 
NGC 4123, as can be seen from the rotation curve 
mass models in Figure 8 of Paper I.  Since we compare the gas velocities 
to data only within the bar, the lack of gas self-gravity 
has little effect on the results.

We use a grid having 256 by 512 cells, each 84.7 pc square, and enforce a 
180\deg\ rotation symmetry so that the grid is effectively 512 by 512 and 
bisymmetric.  Tests show that the results are not dependent on the 
cell size.
The barred potential rotates at a fixed pattern speed \omegap, and 
the grid rotates with the bar.  The $x$-axis of the grid is along the line 
of nodes of the galaxy disk.  This puts the major axis of the bar at an 
angle of 53\deg\ to the $x$-axis.  (It is desirable to avoid having the 
bar aligned with either the $x$- or $y$-axis of the grid.)  The simulation 
timestep is variable, chosen automatically via a Courant condition, and is 
generally 0.1 -- 0.2 Myr.  

We take the sound speed of the gas to be 8 \kms\
(similar to the Galactic value, \eg\ Gunn, Knapp \& Tremaine 1979), 
corresponding to a temperature of $10^4$~K, a preferred temperature
for the diffuse interstellar medium.  
Varying the sound speed within reasonable limits of several \kms\ 
does not materially affect the derived gas flow.  The strength and
shape of the velocity jump across the primary shocks 
(see Figure~\ref{fig-cutml}) do not change much for simulations
with sound speed $c_s = 1 $ to $\sim 25$ \kms.  For $c_s \gtrsim 50$ \kms,
the magnitude of the velocity jump decreases: the sound speed
becomes comparable to the streaming motions, and the shocks are 
no longer strong since the Mach number is not much greater than 1.
However, such high sound speeds for the ISM are unrealistic;
additionally, high sound speeds would require a more massive bar
to produce the observed shock, and the bar we infer is already
near-maximal (Section \ref{sec-comparison}).

\beginfigtwo[t]
\begin{center}
\includegraphics[width=5.05truein]{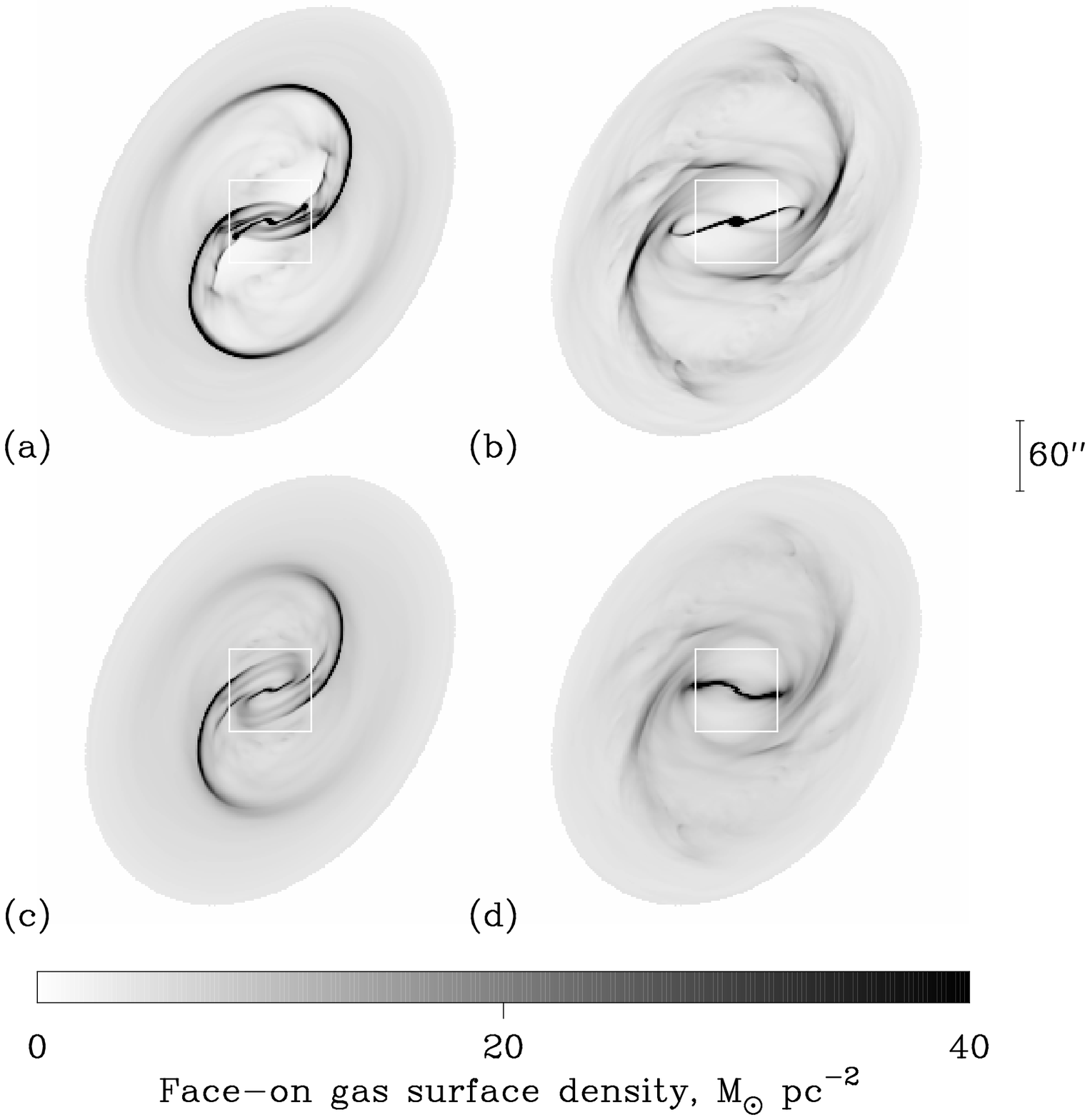}
\end{center}

\caption{Projected view of the gas density field in four models} 
\captsize
The gas density field in four representative
simulations,
at 0.2 Gyr simulation time, projected to the galaxy's orientation on the 
plane of the sky; north is up and west is to the right. The grayscale runs 
from 0 to 40 \msurfden\ in face-on gas surface density.  The bar runs
east-west, and the offset regions of high gas density along the bar
are the loci of shocks.  The white box indicates the bar region 
plotted in Figures \ref{fig-vfields} and \ref{fig-resfields}.
(a) Heavy disk, fast bar model with 
$\ml = 2.25, \omegap = 20~\kmskpc$.  
(b) Heavy disk, slow bar model with 
$\ml = 2.25, \omegap = 10~\kmskpc$.  
(c) Light disk, fast bar model with 
$\ml = 1.0, \omegap = 20~\kmskpc$.  
(d) Light disk, slow bar model with 
$\ml = 1.0, \omegap = 10~\kmskpc$. 
\label{fig-gasdens}
\endfigtwo

\subsection{Initial conditions}

We begin each simulation with an axisymmetric potential so that the gas 
starts on circular orbits.  To achieve this, we redistribute the only 
non-axisymmetric part of the mass, the stellar disk and bar, by averaging 
it over azimuthal angle.  We then cause linear growth of the bar
by interpolating the gravitational field between the
initial axisymmetric and the final observed barred shape.  We allow 0.1 
Gyr to reach the fully barred state, a time much longer than the timestep 
so that the flow adjusts steadily to the growth of the bar.  The bar 
growth time is roughly one-third of an orbital period for material in the 
bar.

We set the initial gas surface density to be constant at 10 \msurfden\ 
inside a radius of 8 kpc; outside that radius it falls off exponentially 
with a scalelength of 13 kpc.  These values were chosen by examining the 
\hi\ data.  (We did not use an initial density distribution with a central 
hole, because the \hi\ hole is probably caused by the bar once it forms,
and may be filled in by molecular gas.)
Longer bar growth times and different initial density distributions have 
very little effect on the results.

The gas response can never reach a completely steady state, since the gas
inside corotation will always lose angular momentum
to the bar; however, after the bar 
growth time, the evolution of the gas flow pattern is quite 
slow.\footnote{Gas builds up in the center due to torque from the bar, 
which can be significant if the code is run for many rotation periods, 
\eg\ several Gyr.  We turned off gas recycling in the code: though it
decreases mass buildup, it can provoke numerical instability (see
Weiner \& Sellwood 1999).}
We continue our simulations to 0.2 Gyr to allow the gas flow 
to settle after the bar has grown, and to 0.3 Gyr to check that the 
flow has stabilized. 

There are some small changes from 0.1 Gyr to 0.2 Gyr in the gas density 
and velocity fields in the bar region, as the gas response to the fully 
grown bar stabilizes. After 0.2 Gyr, gas continues to fall to the center, 
but there is little change in the velocity field in the bar region from 
0.2 to 0.3 Gyr.   We tested a bar growth time of 0.2 Gyr and 
found no significant difference after the gas response had settled.

In order to compare the simulations to the data, for each simulation we 
projected the gas velocity field at 0.2 Gyr into 
line-of-sight velocities, and scaled and rotated it to match the orientation 
of the observed Fabry-Perot velocity field of the bar region as shown in 
Paper I.

\beginfigtwo[ht]
\begin{center}
\includegraphics[width=5.05truein]{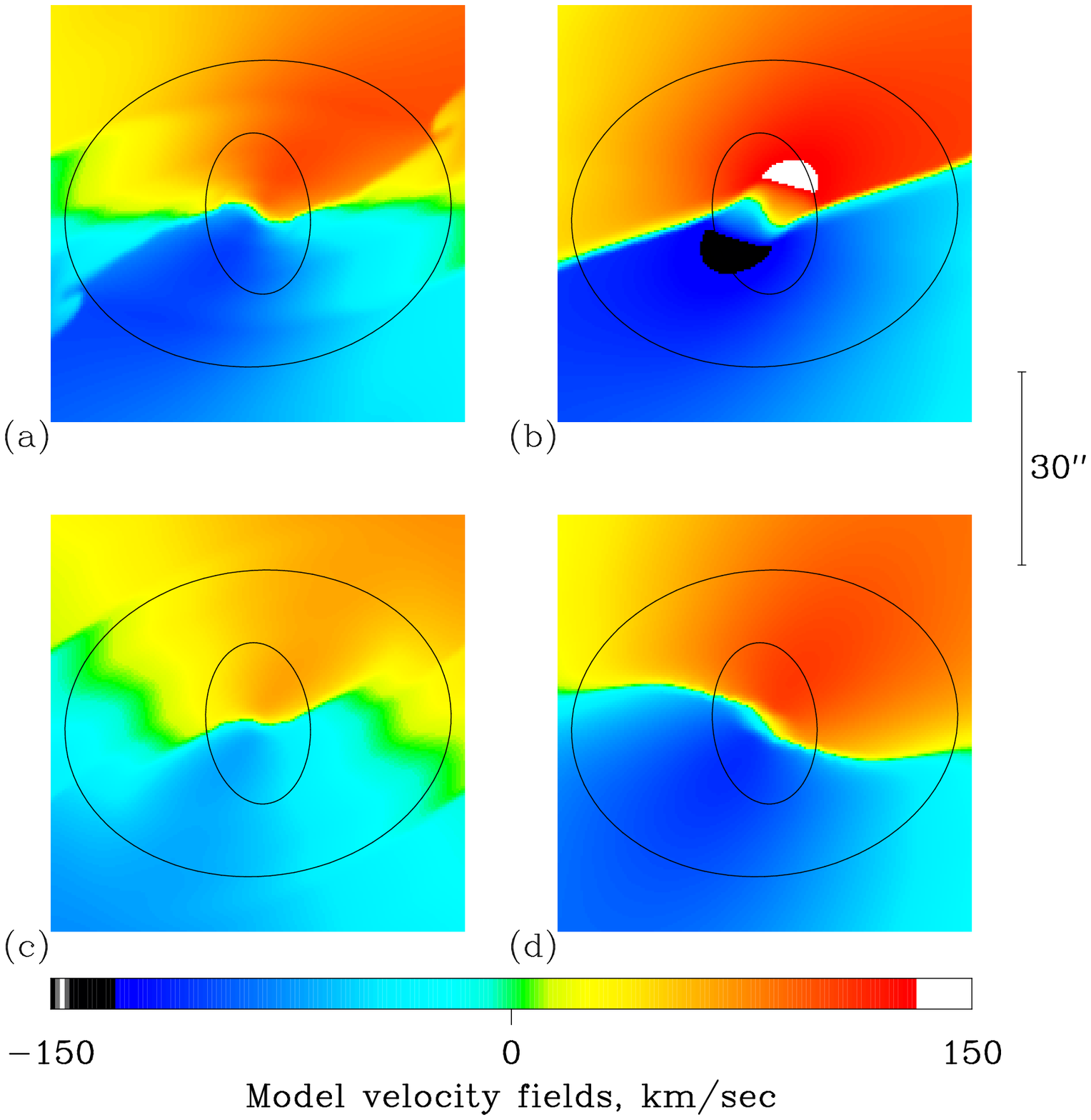}
\end{center}

\caption{Four simulated gas velocity fields}
\captsize
Simulation velocity fields in the bar region
(inset boxes in Figure \ref{fig-gasdens}), 
reprojected onto the plane of the sky to
match the observations. North is up and west is to the 
right.  The projected major axis of the galaxy is at 57\deg\ north of 
west. The color scale shows the line-of-sight velocity field.
The bar runs approximately east-west and 
the shocks are the large velocity gradients perpendicular to the bar.
The area between the ellipses is that used for comparison to
the data in Section \ref{sec-comparison}.  
Panels are as in Figure \ref{fig-gasdens}:
(a) heavy disk, $\ml = 2.25$, fast bar; (b) heavy disk, slow bar;
(c) light disk, $\ml = 1.0$, fast bar; (d) light disk, slow bar.

\label{fig-vfields}
\endfigtwo

\beginfigtwo[ht]
\begin{center}
\includegraphics[width=5.05truein]{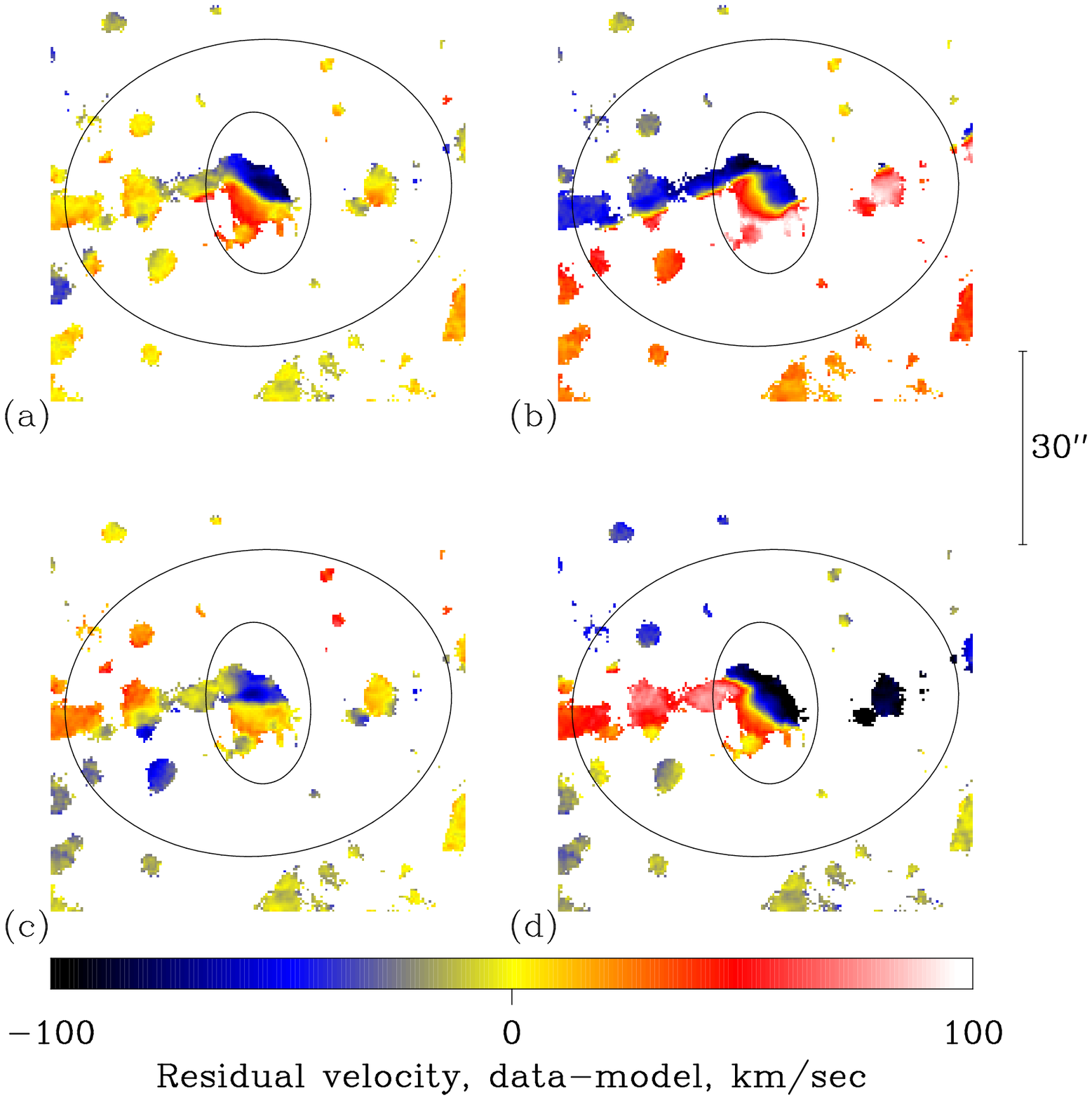}
\end{center}

\caption{Residual velocity fields}
\captsize
Residual velocity fields in the bar region
(inset boxes in Figure \ref{fig-gasdens}),  for four
representative models.  The residuals are the Fabry-Perot
velocity field from Paper I, minus each of the four models
shown in Figure \ref{fig-vfields}.
The orientation is as in Figure \ref{fig-vfields}.
The area between the ellipses is that used for comparison to
the data in Section \ref{sec-comparison}.  All models have
large residuals inside the inner ellipse (8\arcsec\ radius), 
where both data and models are unreliable due to limited
resolution; see Section \ref{sec-compregion}.
Outside that radius, model (a) has significantly smaller residuals 
than the others.  Panels are as in Figure \ref{fig-gasdens}:
(a) heavy disk, $\ml = 2.25$, fast bar; (b) heavy disk, slow bar;
(c) light disk, $\ml = 1.0$, fast bar; (d) light disk, slow bar.

\label{fig-resfields}
\endfigtwo

\subsection{The best fit model and variants}

Figure \ref{fig-gasdens} shows the gas density field viewed at 0.2 Gyr 
run-time in four representative simulations.
Panel (a) has $\ml = 2.25$ and $\omegap=20.0$, 
$\mlratio=0.5$, and an isothermal halo.  This is the best-fitting 
simulation discussed in Section \ref{sec-comparison}.  Panels (b-d)
show variations on this model.  The density fields
have been reprojected and rotated to match NGC 4123's orientation on the 
plane of the sky, for purposes of comparison.  Shocks in the bar, visible 
as regions of high gas density, are associated with jumps in the model gas 
velocity field.  There are multiple regions of high gas density in the 
bar, which correspond to a very strong primary shock and some smaller 
secondary shocks.  The straight features of high gas density traceable 
from the center are the locations of the primary shocks, and correspond to 
the dust lanes visible in the colormap, Figure 2 of Paper I.
The natural explanation for the cause of the shock is the elongation
of the streamlines along the bar (see the discussion of Prendergast 1983). 
Gas falling down along the bar 
potential reaches high velocities, and as the gas climbs
away from the center, up the potential, it decelerates.
Eventually a pile-up occurs and the shock is formed.
The strength and location
of the gradient and shock are strongly influenced by the ellipticity
and rotation rate of the potential.


A massive bar causes a strong spiral pattern in the gas disk outside the 
bar region, as can be seen in Figure \ref{fig-gasdens}.  The spiral 
response is enhanced by the spiral structure and mild ellipticity of the 
stellar disk outside the bar, discussed in Paper I; our simulation method 
causes the outer disk structure to rotate at the same angular speed as the 
bar.  An additional simulation of a model in which we forced the disk 
outside the bar to be axisymmetric still generated a weaker spiral arm pattern.

The spiral pattern continues to evolve from 0.2 Gyr to 0.3 Gyr, propagating 
outward and winding up to tighter pitch angles.  We ran another test 
simulation for a long time (over 1~Gyr) and determined that the spiral 
continues to wind slowly.  The spiral pattern is expected to
evolve on a much longer timescale than the gas flow in the bar since
the orbital timescale (the rotation period) is considerably longer at the 
larger radii where the spiral pattern occurs, and forcing from the bar is 
greatly reduced outside the bar radius.  The steadiness of the gas flow 
pattern inside the bar allows us to compare a snapshot of the simulation 
to the observed velocity field inside the bar radius.  The evolution of the 
spiral pattern renders a snapshot comparison to observations outside the bar 
radius unreliable; fortunately, most non-circular motions are inside the 
bar radius, as discussed in Paper I.


\section{Features of the simulated velocity fields}
\label{sec-vfields}

The models consist of two-dimensional velocity fields over the 
two-dimensional (\ml,\omegap) parameter space with the two subsidiary 
dimensions of \mlratio\ and halo type.  It is impossible to show anything 
more than a small fraction of the models without overwhelming the reader.  
We present four representative models in Figures \ref{fig-gasdens},
\ref{fig-vfields}, and \ref{fig-resfields}
and discuss some of their features.  These models
are drawn from the set with $\mlratio=0.5$ and isothermal halos, 
and represent both low and high mass disks and low and high pattern speeds.
In Section \ref{subsec-slit}, we also discuss the systematic changes
of the model velocity fields with \ml\ and \omegap, plotting
1-D subsets of the models against the observations.

\subsection{Low versus high mass disks}

Panels (a) and (c) of Figure \ref{fig-vfields} show the effect of varying 
the disk mass.  These are two simulated velocity fields which differ only in 
disk \ml.  The field in panel (a) is for an approximately maximum disk 
($\ml = 2.25$), while the field in panel (c) is for a low-mass disk, with 
$\ml = 1.0$, or about 40\% of the maximum disk.  
Both are for 
fast-rotating bars, with $\omegap = 20$, corresponding to $\rlaga = 1.16$ 
and 1.35, respectively.  The velocity fields have been reprojected into the 
plane of the sky, and rotated to match the observations shown in Paper I.  
The projected major axis of the galaxy is at 57\deg\ north of west, while 
the major axis of the bar runs nearly east-west.

The shocks in the bar are the roughly horizontal (east-west) isovels
in the inner 60\arcsec\ of each simulation.  Across these isovels,
perpendicular to the bar, there is a velocity gradient, particularly
large in panels (a) and (b).
The low \ml\ simulations produce less of a shock in the bar than the 
high \ml\ simulations, as can be seen by examining the bar region of the 
simulated velocity fields in panels (c) and (a).  Strongly 
non-axisymmetric motion in the low mass disk (c) is restricted to the inner 
15\arcsec\ radius ($\sim 1.6$ kpc), and the magnitude of the velocity 
jump across the shock is not large except very close to the bar center.  
By contrast, the heavy disk produces offset shocks parallel to the bar 
which extend to two or three times farther out, and have larger velocity 
jumps.  

The residual velocity fields in panels (a) and (c) of Figure 
\ref{fig-resfields} show the importance of the change in shock
strength with \ml.  The fast bar, high \ml\ simulation in panel (a)
matches the observed velocities fairly
well outside the the ILR.  By contrast, the fast bar, low \ml\
simulation in panel (c) produces large systematic residuals
up to 50 \kms,
north and south of the bar, because the velocity jump across the 
shock is not large enough, as discussed further in Section
\ref{subsec-slit}.  Inside the ILR, all models show large residuals,
which are not meaningful (see Section \ref{sec-compregion});
both data and models are affected by limited resolution,
and the structure of the ILR is more sensitive to \mlratio\
than global \ml.

As discussed above in Section \ref{sec-fluidmod}, both simulations produce 
a spiral pattern driven by the bar; the pattern is stronger in the high 
\ml\ simulation, as is to be expected.  The curving shocks to the 
southeast and northwest are along the spiral arms. The spiral arms in the 
high \ml\ velocity field of panel (a) are in approximately the same 
position as the actual spiral arms observed in NGC 4123 just outside the 
bar.  These spiral arms are also visible in the Fabry-Perot \ha\ velocity 
field of NGC 4123, Figure 4 of Paper I, as chains of large bright \hii\ 
regions.  However, the spiral arms are not a particularly good argument 
for or against any model because the spiral pattern evolves over time, as 
mentioned in Section \ref{sec-fluidmod}.  Furthermore, the strength of the 
spiral arm density contrast and the associated shock are quite strong 
early in their evolution, stronger than is observed; and the observed
spiral pattern actually appears to have a slower pattern speed
than the bar, as discussed in Paper I.  These factors suggest that
the observed spiral pattern in the outer disk is decoupled or at best
in resonance with the bar, rather than directly driven by it
(see Sellwood \& Sparke 1988).



\subsection{Fast versus slow bars}

Panels (a) and (b) of Figure \ref{fig-vfields} show the effect of varying 
the bar rotation speed while holding the \ml\ fixed at 2.25, close to the 
maximum disk value.  Panel (a) is for a fast-rotating bar, with $\omegap = 
20~\kmskpc$ and $\rlaga = 1.35$, while panel (b) is for a slow bar, with 
$\omegap = 10$ and $\rlaga = 2.53$.

The simulation with a fast-rotating bar produces offset shocks which lie 
almost parallel to the bar approximately at the location of the observed 
shocks and their associated dust lanes.  The slow-rotating bar also produces 
shocks, but the magnitude and extent of the shocks are larger, and the 
location of the shocks shifts somewhat.  Both effects occur 
because the gas overtakes the slow-rotating bar at a much higher relative 
speed; see Section \ref{subsec-slit} below.

Panels (a) and (b) of Figure \ref{fig-resfields} show the 
velocity residuals for these models.  While the fast bar, high
\ml\ model does fairly well (outside the ILR), the slow bar model
in panel (b) has very large residuals.  The velocity jump
across the shock is too large and too sharp, causing substantial
residuals of $\gtrsim 50$ \kms; see also Section \ref{subsec-slit}.

Finally, panel (d) of Figure \ref{fig-vfields} shows the velocity field 
for a light disk, $\ml = 1.0$, and slow bar, $\omegap = 10$ ($\rlaga = 
2.38$).  There is a fairly strong shock in this model, despite the low 
\ml, because the gas streams through the bar at higher relative speed.
However, the shock is displaced 
very far upstream (in the direction of galactic rotation) 
compared with the other models; 
it is so far from the bar major axis that it is practically out 
of the bar, in strong disagreement with the observed location.
This disagreement is reflected in the very large residuals,
$\sim 100$ \kms, seen in Figure \ref{fig-resfields}(d).

\section{Comparison of simulated velocity fields to data}
\label{sec-comparison}

We have simulated gas velocity fields for six sets of 88 combinations of 
mass-to-light ratio and pattern speed.  We reproject and rotate the velocity 
field of each simulation into the orientation of the galaxy on the plane of 
the sky, and estimate how well it resembles our observed velocity field.  As 
our goodness-of-fit tests yield remarkably tight constraints on our main 
parameters, \ml\ and \omegap, we also offer simple physical explanations for 
the systematic changes, which we illustrate by taking one-dimensional cuts 
through the velocity fields for several ranges of models.

\subsection{The region used for the comparison}
\label{sec-compregion}

For a meaningful comparison of models to data, it is necessary
to use only regions where models and data are primarily
sensitive to the parameters being varied and not to other systematic
effects.
We compare the models and data only in the area between the two ellipses 
on the sky indicated in Figures \ref{fig-vfields} and \ref{fig-resfields},
and in Figure 4 of Paper I.  The outer has diameters 
60\arcsec\ by 48\arcsec, the inner has diameters 24\arcsec\ by 16\arcsec.  
The outer ellipse is chosen to include nearly all the emission within the 
bar, but exclude the large \hii\ regions at the end of the bar, near the 
beginning of the spiral arms.  The inner ellipse is chosen to exclude the 
knot of emission at the very center of the bar.  

We chose this region of the data for several reasons.
The region of most interest is inside the bar radius, where the 
non-axisymmetric motions are strong.  Outside the bar radius, there are 
plenty of velocities with small errors derived from strong \ha\ emission, 
but including them in the comparison would place a large statistical 
weight on regions where the bar-induced motions are weak and therefore 
carry little useful information.  Furthermore, the spiral arm patterns, 
which are the main source of non-circular motions outside the bar, evolve 
over time (Section \ref{sec-fluidmod}).  While the galaxy itself does have 
spiral arms and associated velocity perturbations, comparing them to the 
spirals in simulations which have not settled would be misguided.

We also exclude data at the very center of the galaxy.  The velocity 
gradients are extremely high in this region, and are probably unresolved 
in our data.  Also, there are several reasons why our models are most 
unreliable in this region: our model for the nucleus is probably too soft, 
deprojection errors from the thicker part of the bar are likely to be the 
most significant there, our assumption of uniform $z$-thickness 
is likely to be violated, 
and the resolution of the gas-dynamical code is also worst in the 
center, in the sense of having relatively few grid cells across the scale 
of a streamline.  None of these problems are troublesome 
outside $R \sim 10\arcsec$, but together they force us to exclude data 
within that radius from the comparison.

The distribution of \ha\ emission, and hence velocity data, is
rather sparse within our comparison region.  However, there is 
still a substantial amount of data.  In particular, there are
emission regions within and on either side of the bar,
which allow us to measure the strength and the location
of the main bar shock.  As seen in Figure \ref{fig-resfields},
all of the emission regions within the bar (but outside the ILR)
provide power to discriminate between models.

\subsection{Overall comparison and likelihood}

Within the comparison 
area -- including the bar, but excluding the central region and 
the disk outside the bar -- we can simply use our observed 
velocities and their associated errors to compute the chi-squared 
statistic as a measure of how well the models fit the data.  Since our 
velocity data are oversampled, on 0.36\arcsec\ pixels but with 1.3\arcsec\ 
seeing, we binned down the velocity with a $2\times2$ bin, to provide 
0.72\arcsec\ sampling as a compromise between preserving resolution in the 
velocity field and attaining statistical independence for adjacent pixels. 
This sampling approximately matches the models, which we do not need to 
smooth to equalize their resolution.   Other choices, such as no binning, 
or a $3\times3$ bin (which undersamples the seeing) produce different 
values of reduced \chisq, but the relative quality of the models is 
similar to that presented below.  Overlarge binning makes \chisq\ 
artificially low, since it smooths both data and model heavily.

\beginfig[t]
\begin{center}
\includegraphics[angle=-90,width=3.5truein]{fig4.ps}
\end{center}

\caption{Quality of models' fit to the velocity data. I}
\captsize
A contour plot of \rchisq\
over the \omegap\ -- \ml\ plane for the 88 
models in the set with isothermal halos and $\mlratio = 0.5$,
calculated using strictly statistical velocity 
errors.
The contours are $\Delta(\rchisq)$ = 0.1, 0.2, 0.5, 1, 2, 5, 10, 20, 50,
100, with $N=471$, surrounding the minimum: $\rchisq = 3.54$,
at $\ml=2.25$ and $\omegap=20$.

\label{fig-chisq0}
\endfig

After the $2\times2$ binning, we have $N=471$ velocity measurements within 
the region of interest, with errors ranging from 2.5 to 12 \kms; the median 
error is 7.4 \kms.  These errors are strictly statistical errors from the 
Voigt profile fitting to the Fabry-Perot data (see Paper I).  For each 
simulation, we calculated \chisq\ of the model, 
after computing the best-fit galaxy systemic velocity
and subtracting it from the observations.  For the set 
with isothermal halos and $\mlratio=0.5$, the reduced value \rchisq\ is 
tabulated with the models in Column 4 of Table \ref{table-params}.  
Examining the results in Table \ref{table-params} and 
Figure \ref{fig-chisq0}, it is clear that heavy-disk, fast-bar models with 
$\ml \geq 2.0$ and $\omegap \geq 18$ are the best match to the data.  The 
best fit is for $\ml=2.25$ and $\omegap = 20$.  Light disks and slow bars 
yield miserable fits.

The best models are not perfect fits to the data, in the sense of attaining 
$\rchisq \le 1.0$.  Given the number of idealizations inherent in the 
modeling procedure, this is not surprising, and we cannot expect the 
models to reproduce every feature of the data.

\beginfig[t]
\begin{center}
\includegraphics[angle=-90,width=3.5truein]{fig5.ps}
\end{center}

\caption{Quality of models' fit to the velocity data.  II} 
\captsize
A contour plot of \rchisq\
over the \omegap\ -- \ml\ plane for the 88 
models in the set with isothermal halos and $\mlratio = 0.5$,
calculated using the statistical 
velocity errors and an additional 8~\kms\ dispersion.
The contours are $\Delta(\rchisq)$ = 0.1, 0.2, 0.5, 1, 2, 5, 10, 20, 50,
with $N=471$, surrounding the minimum: $\rchisq= 1.40$,
at $\ml=2.25$ and $\omegap=20$.
\label{fig-chisq8}
\endfig

Another issue is that the purely statistical errors on the velocity data 
are not completely realistic.  Many pixels in bright \hii\ regions have 
errors much less than 8~\kms\ which are highly weighted in the computation of 
\chisq, probably excessively so.  
A given 
line of sight is likely pass through one to a few \hii\ regions, each of 
which may have a velocity different from the bulk matter-weighted 
velocity; our measurement is emission-weighted, and biased toward dense 
regions, since \ha\ emission measure is proportional to $n_e^2 l$.
Hence our \ha\ measurements may not sample the mean velocity field
fairly, and the velocity uncertainties should be adjusted for the 
rms offset of an individual \hii\ region from the mean.

In order to get a sense of the importance of this effect, we added an 
additional {\it ad hoc} dispersion component of 8~\kms\ in quadrature to 
the initial error estimate.  This is equal to the gas sound speed we 
assumed in our models and a reasonable estimate for the velocity 
dispersion among \hii\ regions.  The effect is to reduce the weight of 
very-low-error velocity measurements somewhat, giving the ``better'' 
values of \rchisq\ tabulated in Column 5 of Table \ref{table-params}.  As 
a trade-off, some discriminatory power is lost -- the difference in 
goodness-of-fit between models is somewhat lessened.  However, the 
conclusions are unchanged; light disks and slow bars are still grossly 
worse fits to the data.  In practice, the purely statistical velocity 
errors are certainly underestimates, since no systematic effects have been 
budgeted for, while the boosted error bars probably make the models look 
too good.  The two values of \rchisq\ in Columns 4 and 5 of
Table \ref{table-params} then bracket the 
amount of confidence one should have in a given model.

Figures \ref{fig-chisq0} and \ref{fig-chisq8} show contour plots of 
\rchisq\ distributed on the \omegap\ -- \ml\ plane.  Figure 
\ref{fig-chisq0} shows the values of \rchisq\
calculated using the strictly statistical velocity error bars, and 
Figure \ref{fig-chisq8} shows \rchisq\
calculated with the additional 8~\kms\ dispersion added to the error budget. 
These figures show graphically -- in both senses of the word -- that only 
heavy-disk, fast-bar models work.  Low \ml\ disks and slow bars are ruled 
out.


The models which fit the data reasonably well range from $\ml = 2.0$ to 
2.5, and from $\omegap = 18 $ to 24 \kmskpc\
($\rlaga = 1.5$ to 0.8).  The disk of NGC 4123 is 
between 80\% and 100\% of the maximum disk mass and the bar is 
fast-rotating.

\beginfigtwo[ht]
\begin{center}
\includegraphics[width=4.0truein]{fig6.ps}
\end{center}

\caption{Quality of models' fit to the velocity data, for all models} 
\captsize
The value of \rchisq\ is shown 
over the \omegap\ -- \ml\ plane, for all six sets of
models, calculated using strictly statistical velocity errors.  
Each set contains 88 models.  The left column is for isothermal
halos, the right for pseudo-NFW halos.  The top row has $\mlratio=1$,
the middle row $\mlratio=0.5$, the bottom row $\mlratio=0.0$.  The
contours are $\Delta(\rchisq) = 0.1,0.2,0.5,1,2,5,10,20,50,100$.  \\ 
(a) Isothermal halos, $\mlratio = 1.0$, best $\rchisq=4.90$.
(b) NFW halos, $\mlratio = 1.0$, best $\rchisq=5.02$.
(c) Isothermal halos, $\mlratio = 0.5$, best $\rchisq=3.54$.
(d) NFW halos, $\mlratio = 0.5$, best $\rchisq=4.08$.
(e) Isothermal halos, $\mlratio = 0.0$, best $\rchisq=3.74$.
(f) NFW halos, $\mlratio = 0.0$, best $\rchisq=3.94$.
\label{fig-sixchisq}
\endfigtwo

If instead of adding an additional 8 \kms\ dispersion error component, we 
add 10.8 \kms\ dispersion, then the best model, with $\ml =2.25$ and 
$\omegap=20$, has $\rchisq = 1.00$.  We can then use the values of 
$\chisq$ obtained using this additional 10.8 \kms\ error to determine 
confidence limits.  Formally, all models but the best few are ruled out.  
At the 99.99\% confidence level, the models allowed must have 
$\Delta\chisq < 20$, or $\Delta(\rchisq) < 0.04$.  
The only models allowed are the two with 
$\ml=2.25$ and $\omegap = 20, 21$.  Even if we attempt to compensate for 
the correlation in measurements introduced by the seeing, assuming the 
number of independent measurements is only one-third of $N=471$, the only 
models allowed at 99.99\% confidence are those with $\ml=2.25$ and 
$\omegap = 20$ to 24, and the model with $\ml=2.5$ and $\omegap=23$.

In practice, these restrictive limits are partly due to the assumption 
that all the sources of error are Gaussian.  The level of 
statistical significance is nonetheless impressive -- models in Figures 
\ref{fig-chisq0} and \ref{fig-chisq8} with somewhat worse values of 
\rchisq\ are in fact much poorer matches to the data.

\subsection{Results for other model sets}

We show in Figure~\ref{fig-sixchisq}
the results for all six sets of models, again 
plotting $\Delta(\rchisq)$ over the \omegap-\ml\ plane for each set.  
Although the 
variation of \rchisq\ for different sets differs slightly in the details, 
the primary result is the same.  The best model always has \ml\ of 2 or 
2.25 and \omegap\ in the range 18--23 \kmskpc.  Models with $\ml < 2$ or 
$\omegap < 18$ are always significantly worse.  Figure~\ref{fig-sixchisq}
shows that the subsidiary
parameters of halo shape and \mlratio\ have little effect on
the distribution of $\Delta(\rchisq)$.

We also computed the results for all sets of models when an additional 
8~\kms\ dispersion is added to the statistical errors on the observed 
velocities, as described above.  This procedure weights the observed 
velocities slightly differently, and in down-weighting velocity 
measurements with extremely small errors, is probably a more realistic 
basis for choosing the best-fit model.  When we compute values of \rchisq\ 
with the added 8~\kms\ dispersion, the best model, with $\ml=2.25, 
\omegap=20, \mlratio=0.5$, and isothermal halo, remains the best; the 
model with the same parameters but $\mlratio=0$ is almost as good.  For 
each set of \mlratio\ and halo type, the best model always has $\ml=2.25$ 
and \omegap\ in the range 20 -- 23 \kmskpc.  Models with $\ml < 2$ and 
$\omegap < 18$ are always ruled out at the 99.99\% confidence level.

No matter how we model the nucleus and the halo, or treat the errors on the 
observed velocities, Figure \ref{fig-sixchisq}
shows that the best models are always in the same region of \ml\ 
and \omegap, requiring a heavy disk and fast bar.
The six sets of models have best values of \rchisq\
which differ somewhat: there is some preference for a low \ml\
for the nucleus, which is plausible since it is blue and an
emission-line source.  There is not a strong
discrimination between NFW and isothermal halo profiles.
Overall, the best value for disk \mli\ is 2.25, and 
$\mli=2.0$ is a strong lower limit.

Since we have compared the models to velocity data only within the bar
radius, technically we require only that \ml\ must be high 
within the bar radius.  However, in Paper I we showed that 
an \ml\ which begins to decline outside the bar radius does not
cause a significant decrease in the disk contribution to
the rotation curve.  Therefore, the high \ml\ indicates
that the disk contribution is at or near maximum.


\subsection{A slit cut through the data and models}
\label{subsec-slit}


The previous comparisons of our models with 2-D data indicate very 
tight constraints on our two main parameters.
In order to gain a physical understanding of these constraints, 
we examine a subset of the 
data in more detail.  As we emphasized in Paper I, the strongest 
signatures of the non-axisymmetric motions caused by the bar are the 
offset shocks along the bar, which are also the locations of the dust 
lanes.  We extract a slice nearly perpendicular to the bar from the 
Fabry-Perot data 
in order to show the velocity jump in the shock, which we compare with the 
corresponding data from the models.
We make this 
comparison for two subsets of the 2-parameter \ml\ -- \omegap\ space 
which include our best-fitting model, for which $\ml = 2.25, \omegap = 20$.

\beginfig[t]
\begin{center}
\includegraphics[width=3.5truein]{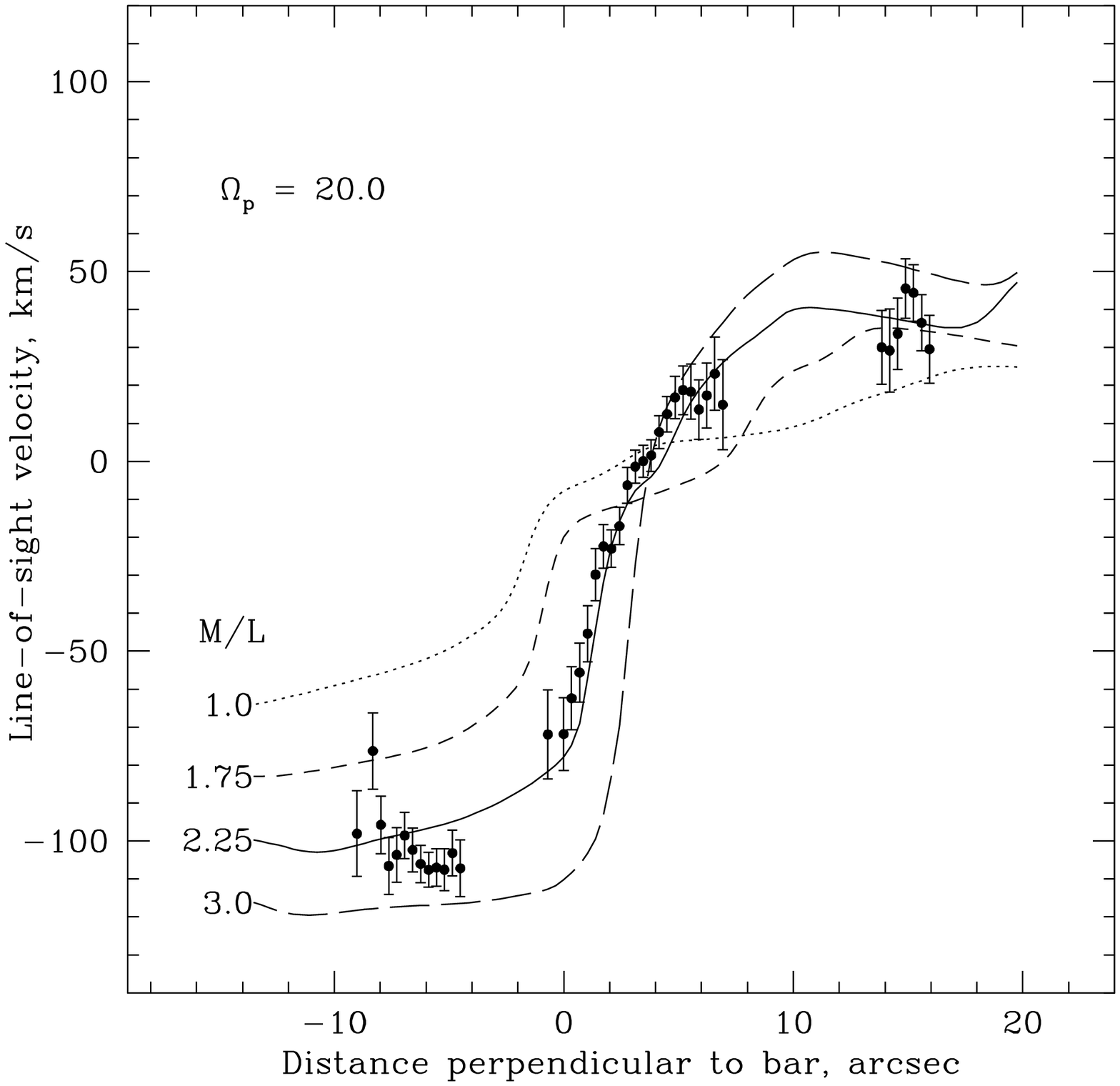}
\end{center}

\caption{A cut through the velocity data and models of varying \ml} 
\captsize
The Fabry-Perot velocity data (points) and associated errors from a 
pseudo-slit cut through the data, nearly perpendicular to the bar.  The 
x-axis is distance from the bar in arcsec, The y-axis is line-of-sight 
velocity. The lines are the velocity fields of models; all have $\omegap = 
20$. The \ml\ values and line styles are 1.0 (dotted), 1.75 (short dash), 
2.25 (solid), and 3.0 (long dash).
\label{fig-cutml}
\endfig

The position of our pseudo-slit through the Fabry-Perot data is indicated 
by the vertical line in Figure 4 of Paper I.  It is at 17\arcsec\ (2 kpc 
distance in the disk) east of the galaxy center, runs north-south, nearly 
perpendicular to the bar, and is 2 pixels, or 0.72\arcsec, wide.  We 
selected it to pass through several regions of \ha\ emission, including 
one which straddles the velocity jump associated with the shock, and two 
regions of \ha\ emission on either side of the shock.
The spatial resolution of the data and models shown here 
are similar -- about 1.4\arcsec\ for both.

The velocity data are the points with error bars plotted in Figures 
\ref{fig-cutml} and \ref{fig-cutomegap}, while the predictions from
sets of models are drawn as smooth curves.  The $x$-axis is 
distance from the bar major axis in arcsec, in the direction of galaxy 
rotation, so that gas overtakes the bar moving from left to right on this 
plot.  The $y$-axis is projected line-of-sight velocity.  The strong 
gradient in the central part of the data is a combination of the changing 
projection of the gas streamlines, and the shock at the leading edge of 
the bar.  The gradient is clearly spatially resolved.  The two \hii\ 
regions away from the shock do not have significant
internal velocity gradients.

The change in velocity from one side of the bar to the other 
is 140~\kms\ in line-of-sight velocity (uncorrected for
inclination), a substantial fraction of the velocity width
of the galaxy.
This gradient arises from the changing projection
of the non-circular streamlines and the shock in the bar
(see the velocity field in Paper I).  
The shock is caused by elongation of the streamlines along the
bar, so its strength is closely dependent on the bar mass,
which controls the ellipticity of the potential.

\subsubsection{Varying the disk mass}

In Figure \ref{fig-cutml}, we keep \omegap\ fixed at 20 and plot four 
models with $\ml = $ 1.0, 1.75, 2.25, and 3.0.  At $\ml = 1.0$, the 
simulation predicts a rather small velocity change, less than half that 
observed, but the velocity change increases and becomes steeper as \ml\ 
rises.  Furthermore, 
there is a systematic shift in the position of the strong shock, where the 
gradient is steepest; the much smaller velocity wiggles are due to 
secondary shocks.   Of these four models, only that with $\ml = 2.25$ 
shows a shock of about the right strength at the location of the observed 
velocity jump. 

The constraint on \ml\ is strikingly tight.   Non-axi\-symmetric 
forces become weaker as \ml\ decreases, as the influence of the bar 
declines and that of the round halo grows.
Models with $\ml < 2.0$ cannot produce a strong enough shock, even at 
2 kpc from the galaxy center.  The presence of shocks with a large
velocity jump along the length of the bar requires that the 
streamlines be quite elliptical, which requires a massive bar.
If the dark matter is dominant in the inner disk, the potential
is rounder and cannot produce the shocks.


\subsubsection{Varying the pattern speed}

In Figure \ref{fig-cutomegap}, we keep \ml\ fixed at 2.25 and plot models 
of varying \omegap.  The models plotted have $\omegap = $ 10, 14, 20, and 
24~\kmskpc, corresponding to $\rlaga = $ 2.53, 1.93, 1.35, and 0.68.

As the pattern speed is decreased and the Lagrange radius increased, the 
shock moves upstream, in the direction of rotation of the galaxy and of 
the bar.  This behavior occurs because at slower bar pattern 
speeds, the gas overtakes the bar more quickly, and climbs farther out of 
the bar potential before shocking.  The higher velocity of the gas in the 
co-rotating frame also causes the shock to steepen for slower bars, as 
noted in Section~\ref{sec-vfields}.  At $\omegap = 10$, the 
pattern speed is so slow that the velocity gradient
of the shock is extremely steep.  This model is
so extreme that the global flow pattern changes
and the relation between pattern speed
and shock location breaks down.

The behavior of models in this pseudo-slit cut through the data shows 
clearly why the observations favor models with high \ml\ and high \omegap. 
This is also generally true of the models not shown in either Figure 
\ref{fig-cutml} or \ref{fig-cutomegap}.  There is some covariance between 
the parameters \ml\ and \omegap, but their effects on the location and 
gradient of the shock are rather different. The best models have high mass 
disks and fast bars, and it is not possible to rescue light mass disks 
with $\ml < 2.0$ by appealing to slow bars.

\beginfig[t]
\begin{center}
\includegraphics[width=3.5truein]{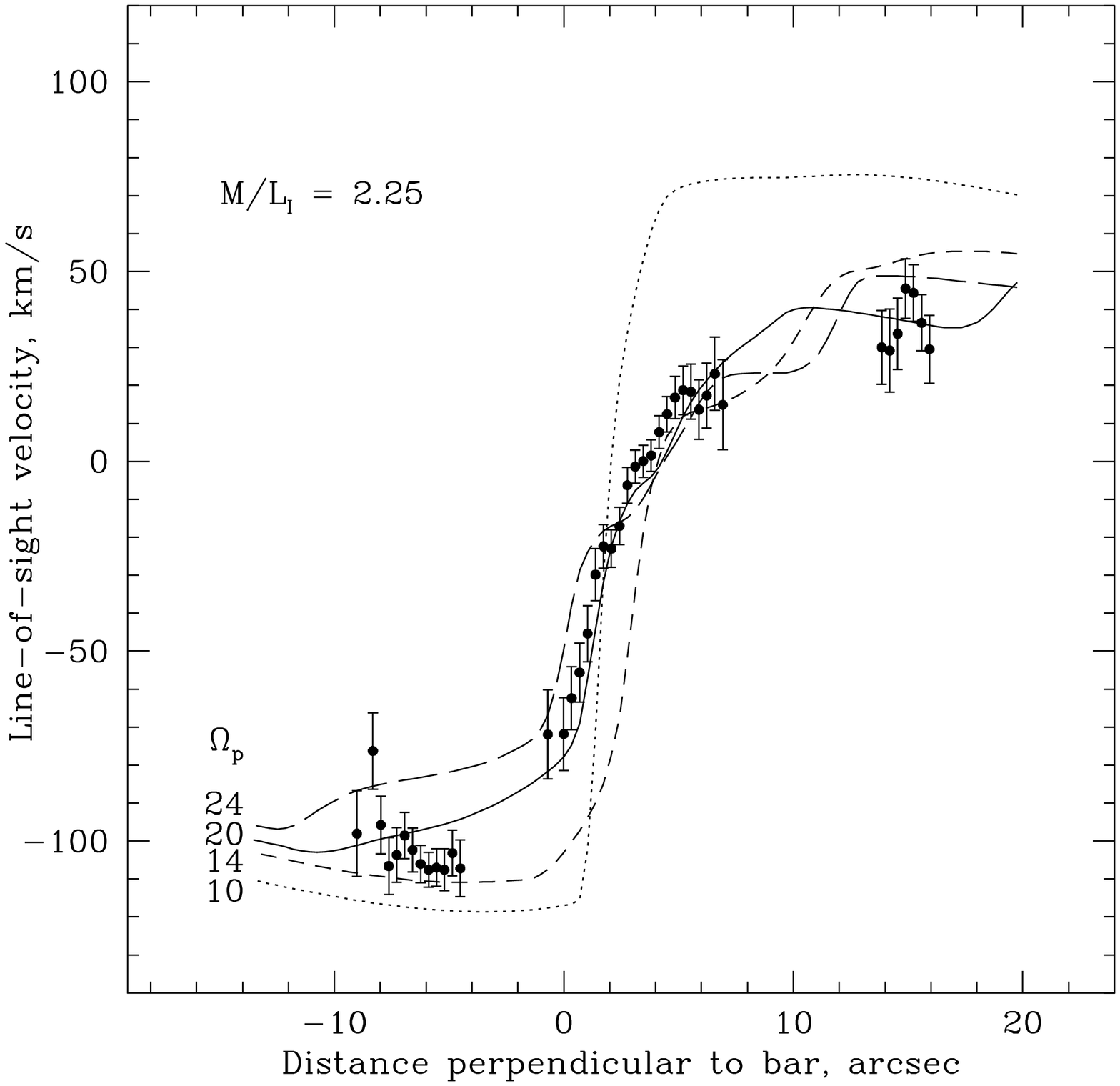}
\end{center}

\caption{A cut through the velocity data and models of varying \rlaga} 
\captsize
The Fabry-Perot velocity data (points) and associated errors from a cut 
through the data, nearly perpendicular to the bar.  The x-axis is distance 
from the bar in arcsec, The y-axis is projected line-of-sight velocity. 
The lines are the velocity fields of models; all have $\ml = 2.25$. The 
\omegap\ values are 10.0 (dotted), 14.0 (short dash), 20.0 (solid), and 
24.0 (long dash).
\label{fig-cutomegap}
\endfig


\section{Discussion I.  Properties of the dark halo}
\label{sec-haloprop}

Our measurement of the disk \ml\ of NGC 4123 determines the
disk mass distribution, and, since we have an extended
\hi\ rotation curve, the mass distribution of the dark halo.
We find that the disk dominates the mass inside the optical
radius $R_{25}$; see the rotation curve decompositions in Paper I
for $\ml=2.25$.  We have only used data inside the bar radius,
but the result is robust: the disk dominates the mass
even if its \ml\ were to decline outside the bar radius
(Section 6.3 of Paper I).
Dark matter halos with either isothermal
profiles with a constant-density core, or power-law profiles
with an inner cusp, are both consistent with the data,
as discussed in Paper I.
However, the halo rotation curve $V_{halo}(R)$ and equivalently
the enclosed mass $M_{halo}(<R)$ are fairly well constrained,
especially at large $R$.  Hence we have detailed knowledge
of the radial distribution of luminous and dark mass in this galaxy.
In this section we use the mass distributions to test 
models of the relation between disks and their dark halos,
and of the process of disk formation by dissipative collapse 
within the halo.

\subsection{Comparison to ``universal'' relations}

The tight correlation between galaxy luminosity and rotation width, the 
Tully-Fisher relation, suggests that there is some kind of universal 
relation between galaxy disks and halos, especially since galaxy rotation 
curves are approximately flat out past the optical radius.  
Recently there 
have been a number of suggestions for universal relations which govern the 
disk-to-halo ratios of galaxies, which can be tested against NGC 4123.
Additionally, models of structure formation yield
predictions for dark matter halo parameters.

The ``Universal Rotation Curve'' claimed by Persic, Salucci \& Stel (1995) 
gives a relation of disk and halo mass with rotation
velocity, and suggests that galaxies of smaller rotation width 
are more dark matter dominated.
For NGC 4123, their relation predicts that dark matter
should become dynamically detectable at $0.6R_{25}$ (6.7 kpc)
and that the mass within this radius is 40\% dark.
In contrast, we find that the galaxy is at most 15\% 
dark matter within that radius.  

Several studies motivated by cosmological simulations of structure 
formation have examined the structure of dark matter halos.
Such simulations produce centrally 
concentrated dark halos, generally with a broken-power law density 
profile, with a central cusp slope of $r^{-1}$ or $r^{-1.5}$
(\eg\ Navarro \etal\ 1996; Syer \& White 1996; 
Kravtsov \etal\ 1998; Moore \etal\ 1999; Klypin \etal\ 2000).
However, these studies do agree that the central concentration
of dark halos is dependent on the mass of the halo and
on the cosmological model.  Such halos follow one-parameter
families: the halo scale density and radius are linked.

It is tempting to identify a one-parameter halo family with
the Tully-Fisher relation and derive a relation between 
halo scale and rotation width.  Navarro \etal\ (1996) proposed a relation 
of halo to galaxy which produces galaxies that are very dark matter 
dominated: a galaxy with a maximum rotation of 200 \kms\ is 90\% dark 
matter within the optical radius, and a galaxy with maximum rotation of 
100 \kms\ is 96\% dark matter within the optical radius.  

The $\Omega = 0.3$ model of Navarro (1998a), where there is less 
dark matter to go around, still produces very dark matter dominated 
galaxies.  Figure 2 of 
Navarro (1998a) proposes a disk-halo decomposition of the well-studied 
galaxy NGC 3198, whose rotation width is about 150 \kms, similar to that 
of NGC 4123.  
In the model of Navarro (1998a), NGC 3198 is 75\% dark matter inside the 
optical radius, the dark halo dominates the rotation curve even at 2 
kpc from the galaxy center, and the disk \ml\ is $\simeq 0.8h_{75}$.
This model is clearly inconsistent with our result for NGC 4123.

In Figure \ref{fig-halocomp} we plot the 
(pre-baryonic-collapse) halo scale density and radius predicted 
for a galaxy of circular velocity $V_c=130~\kms$ from
the models of Navarro (1998b), and the confidence ellipses
for our NFW-type halo fits to the \hi\ rotation curve of
NGC 4123, from Paper I.
In all cases, for disk \ml\ from 2.0 to 2.5 and for all three cosmologies, 
the halos determined by our modeling have lower scale density or radius 
(or both) than the CDM halo predictions. The implication is 
that either the dark halo is of similar size to the prediction but less 
dense by a factor of $\sim 3$, or the dark halo has a smaller break radius
than predicted.  Since the inner part of 
an NFW-type halo has $\rho \sim r^{-1}$, a smaller scale radius means a 
lower halo density at a given physical radius in kpc.
In terms of the NFW halo concentration parameter $c$, we find that
the halo of NGC 4123 has $c=2.1$, while Navarro (1998b) predicted
$c \simeq 5$ for the $\Lambda$CDM cosmology with $\Omega_M = 0.3$.

The study of Bullock \etal\ (1999) uses a large number of
simulated halos to determine the $c - M$ relation and scatter.
Although it does not directly propose disk-halo decompositions,
we can find the concentration expected for the halo of NGC 4123.
In a $\Lambda$CDM cosmology, the halo
has a mass of $M_{vir} = 3 \times 10^{11} M_\odot$
for an NFW-type profile.  Using a modified definition of
the concentration, $c_{vir}$, Bullock \etal\ predict
that such a halo will have $c_{vir} = 16.6$ in a 
$\Lambda$CDM model, and the scatter
is given by ${\rm log}~c_{vir} = 1.22 \pm 0.14$.
The halo of NGC 4123 has 
$c_{vir} = 4.4$, or ${\rm log}~c_{vir} = 0.64$, a
$4\sigma$ deviation, requiring a much less dense halo
than predicted.

\beginfig[t]
\begin{center}
\includegraphics[width=3.5truein]{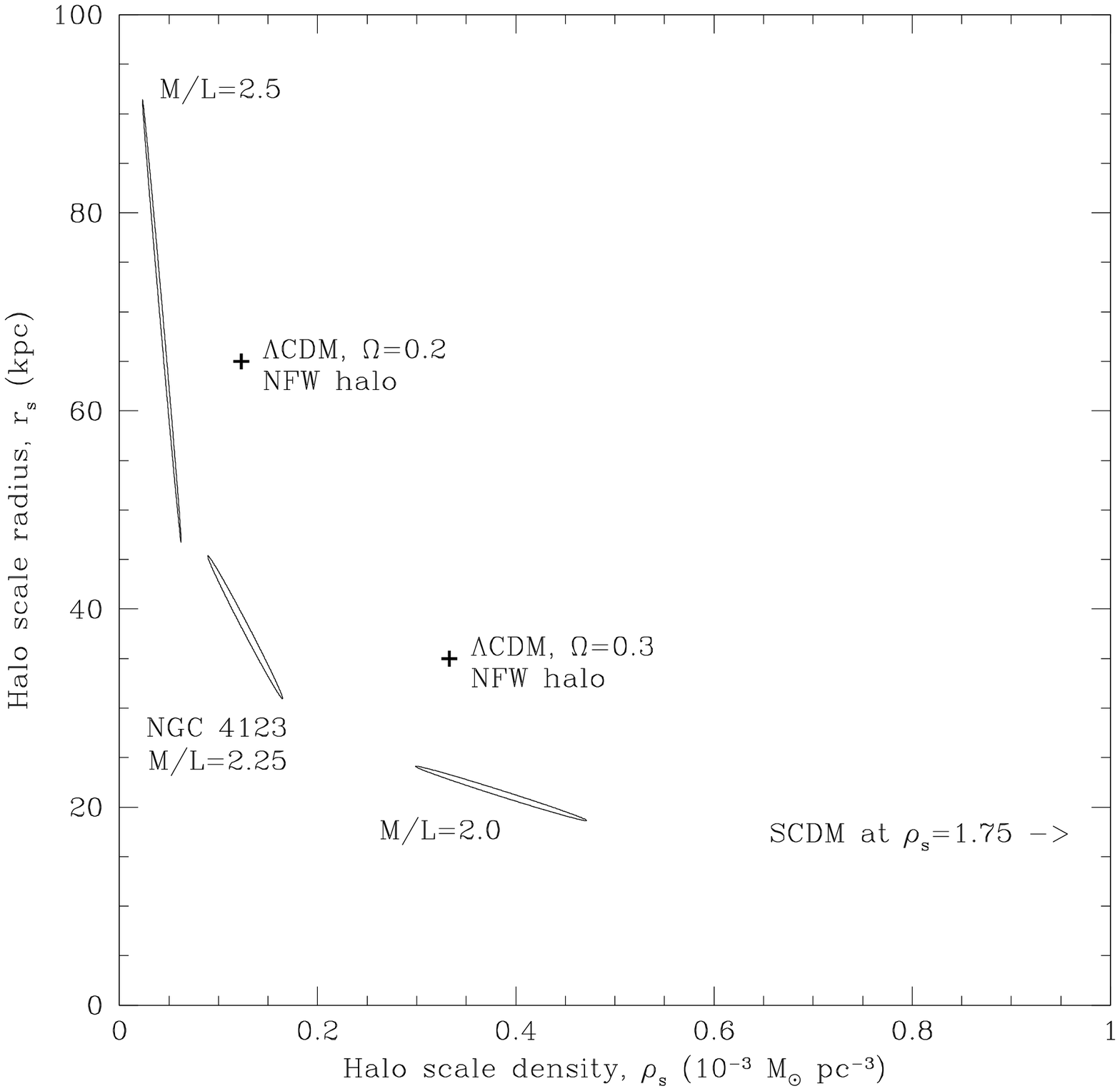}
\end{center}

\caption{Parameters of the dark halo vs. predictions} 
\captsize
The scale density 
$\rho_s$ and radius $r_s$ of the NFW-type dark halos allowed by our models 
compared to the pre-collapse halos predicted by the simulations of 
Navarro (1998b). The $1\sigma$ confidence ellipses in $\rho_s - r_s$ space 
determined from our rotation curve fits in Paper I are shown for the 
allowed disk \ml\ from 2.0 to 2.5. The halo parameters predicted 
for a galaxy of NGC 4123's rotation width are indicated for three 
cosmologies.
\label{fig-halocomp}

\endfig

We emphasize that the dark halo we determine by fitting the rotation curve 
is seen in its final state, after the baryonic disk has collapsed inside 
it. The dissipative baryonic collapse must draw the halo inward
so that the final 
halo has a higher density and smaller radius than the initial state. The 
halo parameter predictions from Navarro (1998b) and Bullock \etal\ (1999)
are pre-collapse halos, which increases the
discrepancy between the predictions 
and our measurements of halo concentration.
In the next section we describe an attempt
to quantify the effects of baryonic collapse.

\subsection{Baryonic collapse and the initial halo density profile}

The dominant paradigm of disk galaxy formation
requires that the baryonic disk collapse within the 
dark matter potential, drawing the halo inward
(White \& Rees 1978; Fall \& Efstathiou 1980; Faber 1981).
The final disk and halo mass distributions, which we have measured
in the form of $V_{disk}(R)$ and $V_{halo}(R)$,
can in principle be used to infer the initial, pre-collapse halo density
profile, which is clearly of great interest.  We show here that 
using the standard prescription to approximate the
collapse process leads to an unrealistic initial profile.
A more sophisticated treatment of the collapse
will be needed to determine the initial density profile
of the dark halo from rotation curve information.

It is generally assumed that collapse into a galaxy
is a two-stage process: first the halo (dark matter and
baryons) forms, then
the baryonic component cools within the dark matter halo,
contracting until it is halted by angular momentum
(Fall \& Efstathiou 1980).  The dissipative baryonic collapse
compresses the dark halo.  The formalism often used
to model this collapse is known as the adiabatic-compression 
or circular-orbit approximation (see \eg\ Faber 1981; 
Blumenthal \etal\ 1986).  This formalism is an integral part 
of many recent treatments of the process of disk formation
(Dalcanton, Spergel \& Summers 1997; Mo, Mao, \& White 1998;
van den Bosch 1998, 2000; Navarro \& Steinmetz 2000), 
which compute the final state of the disk from its 
angular momentum distribution, and use the adiabatic-compression
formalism to compute the post-collapse dark halo mass profile.  

The formalism assumes that (1) the disk 
can be approximated by a sphere with the same 
enclosed-mass profile; (2) the collapse of the disk inside the halo is 
roughly adiabatic; (3) that detailed conservation of angular momentum 
applies; and (4) that ``mass shells don't cross,''
i.e. that dark halo particles behave as if they were all on 
circular orbits.  Under these assumptions, the initial and final halos 
are related by
\begin{equation} r_i M_{i,total}(r_i) = r_f (M_{f,disk}(r_f) + 
M_{f,halo}(r_f))
\label{eq-collapse}
\end{equation} where $M(r)$ indicates the mass interior to $r$, and 
subscripts $i$ and $f$ indicate initial and final quantities.  The halo 
mass shell originating at $r_i$ ends up at $r_f$.  This equation
assumes the conservation of angular action, which holds for
individual dark matter particles, but in order to apply it to 
mass shells one has to assume that mass shells have a well-defined
radius and that they do not cross.  For this reason we prefer
to call the formalism a circular-orbit approximation, although
many authors simply refer to it as ``adiabatic compression'' of
the halo by the disk.

Our determination of the disk \ml, and of the halo parameters from the 
\hi\ rotation curve, give us the final mass profiles of both the disk and 
the halo.  From this it is easy to use Equation \ref{eq-collapse} to find 
the radius $r_i$ corresponding to each $r_f$ (i.e. the collapse factor at 
each radius) and the initial mass profile $M_i(r_i)$, the 
decompressed halo.
We carried this out for models with a
$\ml=2.25$ stellar disk and the corresponding isothermal and NFW-type 
final dark-halo profiles $M_{f,halo}(r_f)$.

The final profiles must be more centrally concentrated than
the initial profiles, so we also fit a very centrally concentrated
final halo to the rotation curve and decompressed that model, using a
``Moore-type'' halo profile with a central density cusp of 
$\rho \propto r^{-1.5}$ (Moore \etal\ 1999).  The parameters
of the fitted final dark halos are given in Table~\ref{table-halos}.
The very centrally concentrated Moore-type halo can make an
acceptable fit to the data, but only at the price of a quite
large scale radius.


\begintab[t]
\begin{center}
\begin{tabular}{lrr}

\tableline\tableline
Halo type  &  Scale density  & Scale radius \\ 
           &    $10^{-3}~\mden$ &  kpc  \\ 
\tableline

Isothermal  &  4.68\phn  &  6.33 \\ 
NFW-type    &  0.127  &  38.2\phn \\
Moore-type  &  0.040  &  61.0\phn \\
\tableline
\end{tabular}

\end{center}
\caption{Halo models: best-fit parameters for $\mli=2.25$}
\label{table-halos}

\endtab

The assumption that mass shells don't cross implies that
$M_{i,halo}(r_i) = M_{f,halo}(r_f)$.
Then the halo can be ``decompressed'' by solving 
Equation~\ref{eq-collapse} for the initial radius.  
Using assumption (1) above, we approximate the disk by
a sphere and convert from $M_{f,disk}$ to $V_d$:
\begin{equation}
\frac{r_i}{r_f} = (1 - f_b) \left(
            1 + \frac{{V_d}^2(r_f)}{{V_h}^2(r_f)} \right)
\label{eq-uncomp}
\end{equation}
where $f_b$ is the baryon fraction, i.e. 
$M_{i,disk} = f_b M_{i,total}$, and
$V_d$ includes the stellar and \hi\ disks.
The effect of the spherical assumption is small (Barnes 1987; Sellwood 1999).
We assumed a baryon fraction of 0.1; different values of
$f_b$ have little effect on the results, since they just
change the scale of $r_i$.

Figure \ref{fig-uncomprad} shows the initial radius as 
a function of final radius for the isothermal, NFW-type, and 
Moore-type halos.  The predominant feature is a break
at $r_f \sim 9$ kpc, which is the location of the peak
in the disk rotation curve (see Paper I).  Inside $R=9$ kpc,
$V_d$ increases steeply, while outside 9 kpc, $V_d$ drops
as $V_h$ rises gently.  Outside 9 kpc, the results for
all models are nearly identical.

At small radii in the isothermal halo model, $r_f$ is not a
monotonic function of $r_i$, which means that the assumption
that mass shells do not cross is not self-consistent.
Under the assumption of the adiabatic circular-orbit
approximation, there is no initial halo profile which
can yield a final isothermal halo profile for NGC 4123.

In the NFW and Moore-type halo models, $r_f(r_i)$ does not
strongly violate monotonicity.  There are a few bumps,
at $r_i=3$, 9, and 20 kpc, which are clearly related to
small-scale features in the stellar+gas rotation curve
(see Figure 8 of Paper I) and do not represent real
features in the collapse profile of the dark halo.

\beginfig[t]
\begin{center}
\includegraphics[width=3.5truein]{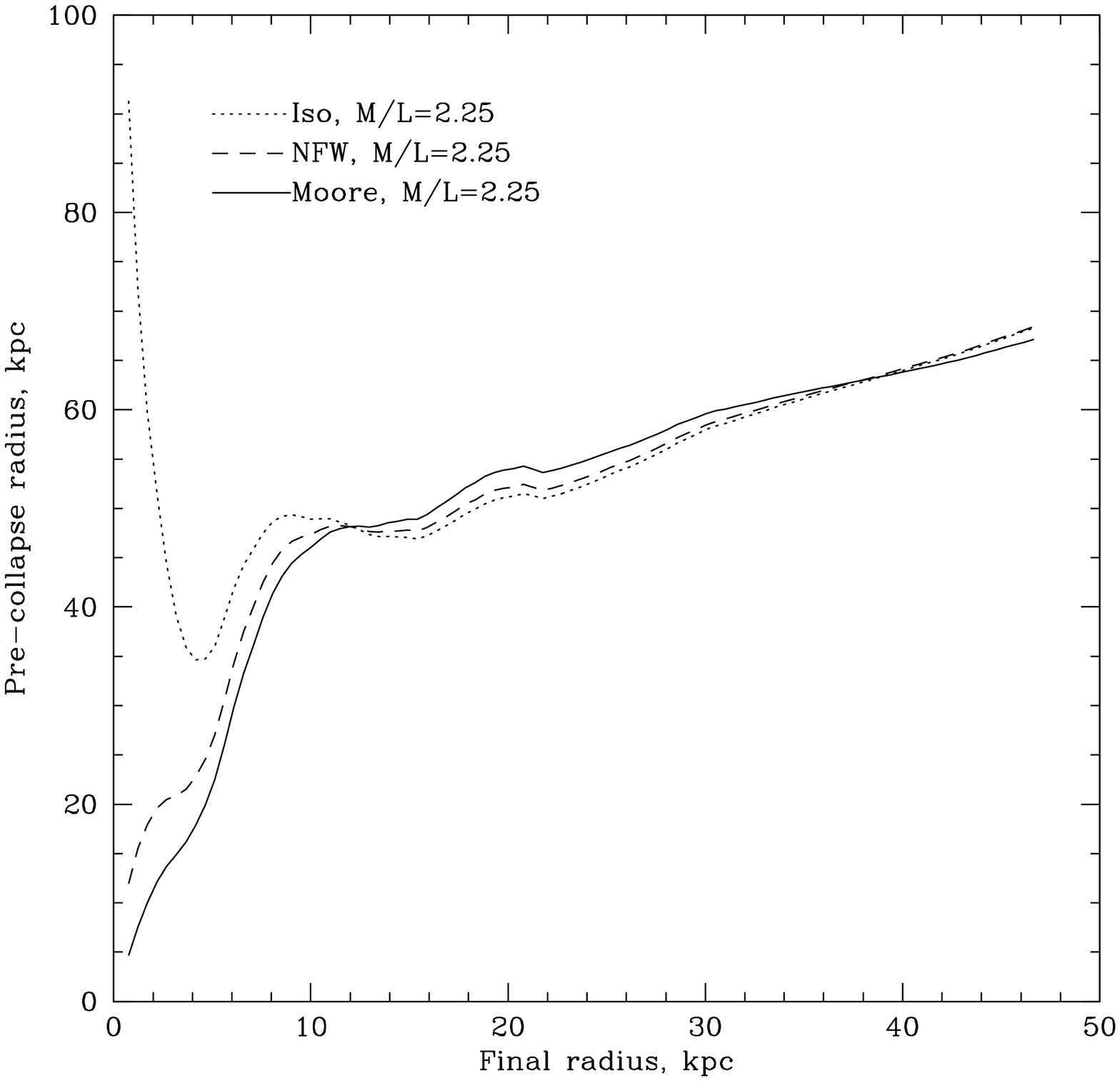}
\end{center}

\caption{Halo compression in the circular-orbit approximation}
\captsize
The initial, pre-collapse radius $r_i$ of a halo mass shell,
obtained using the circular-orbit formalism to uncompress the halo,
plotted against its final radius $r_f$.
The three models have final halo profiles of isothermal 
(dotted), NFW-type (dashed), and Moore-type (solid).
\label{fig-uncomprad}

\endfig

The upper panel of Figure \ref{fig-uncompmass} 
shows the final enclosed-mass profiles of the three
halo models, determined by fitting the rotation
curve.  Since all three models are determined by
the same data, they are very similar, except in the
central 10 kpc, where the rotation curve is dominated
by the stellar disk and the halo is not well constrained.

The lower panel of Figure \ref{fig-uncompmass} 
shows the enclosed-mass profiles
of the initial, uncompressed halos derived from the three
halo models. 
The initial halos are
much less concentrated than the final halos.
For the final isothermal halo, the initial mass profile is 
unphysical, because $r_i(r_f)$ is not monotonic.  
For the final NFW and Moore-type profiles, the initial
profiles are essentially physical, with small 
deviations due to the wiggles in $r_i(r_f)$ caused
by structure in the disk+gas rotation curve.

Outside $r_i \sim 47$ kpc, the mass profiles of
all three models are similar, even for the 
final isothermal halo.
Figure~\ref{fig-uncomprad} shows that this radius
corresponds to $r_f = 9-10$ kpc, the location of
the peak of the stellar+gas rotation curve.
Inside this radius, the halo profile is not
strongly constrained, and the uncompression
is influenced by features in the stellar
distribution.  Outside this radius, the halo
profile is fairly well constrained by the need
to fit the \hi\ rotation curve; the fitted
final mass profiles for isothermal,
NFW, and Moore profiles are similar; and the initial mass
profiles in the three models are also similar.

The derived initial mass profile outside $r_i=50$ kpc is
well described by a power law: $M_{i,halo} \propto {r_i}^{3.3}$,
implying an initial density profile of $\rho_i \propto r_i^{0.3}$,
extending out to $r_i \simeq 70$ kpc.
Inside 50 kpc, this power law also fits the initial profile
for the NFW model; the Moore model is shallower while the
isothermal model becomes unphysical.

A $\rho \propto r^{0.3}$ density profile for the pre-galactic 
initial overdensity is extremely unattractive.  
The problem exists
in all three halo models; it is simply a consequence
of the relative disk and halo rotation curves and
Equation~\ref{eq-uncomp}.  The initial mass profile
is actually best determined at large radius, while a
roughly constant density would only be believable in \eg\ the
inner core of the pre-galactic halo.

We believe that the problem is in the circular-orbit formalism, which can
exaggerate the amount of halo collapse caused by a high mass disk
(Barnes 1987; Sellwood 1999).  The assumption most likely
to be violated is that mass shells do not cross, since a real
halo will have some fraction of radial orbits; radial orbits can provide 
extra stiffness against compression (Barnes 1987).  It is
also possible that detailed conservation of angular momentum
is violated, \eg\ through angular momentum transfer within or
between disk and halo, or through mass loss.  However, 
transfer of angular momentum from disk to halo would
exacerbate the overcooling problem found in CDM+gas simulations,
which results in disks that are too small (\eg\ 
Navarro \& Steinmetz 1997, 2000).

In order to produce the relatively low
density and large scale radius final halos that we obtained from
fitting the rotation curve, the circular-orbit approximation
forces an unrealistically diffuse initial condition.
The use of the circular-orbit formalism to calculate properties of
observed galaxies could be misleading.
It should be possible in future work to derive the initial halo 
properties from measured final disk and halo mass profiles, 
through either $N$-body simulation or analytic models with 
more complex dynamics (\eg\ a model incorporating a
halo distribution function, and in which both angular and 
radial actions are conserved during compression).

\beginfig[t]
\begin{center}
\includegraphics[width=3.5truein]{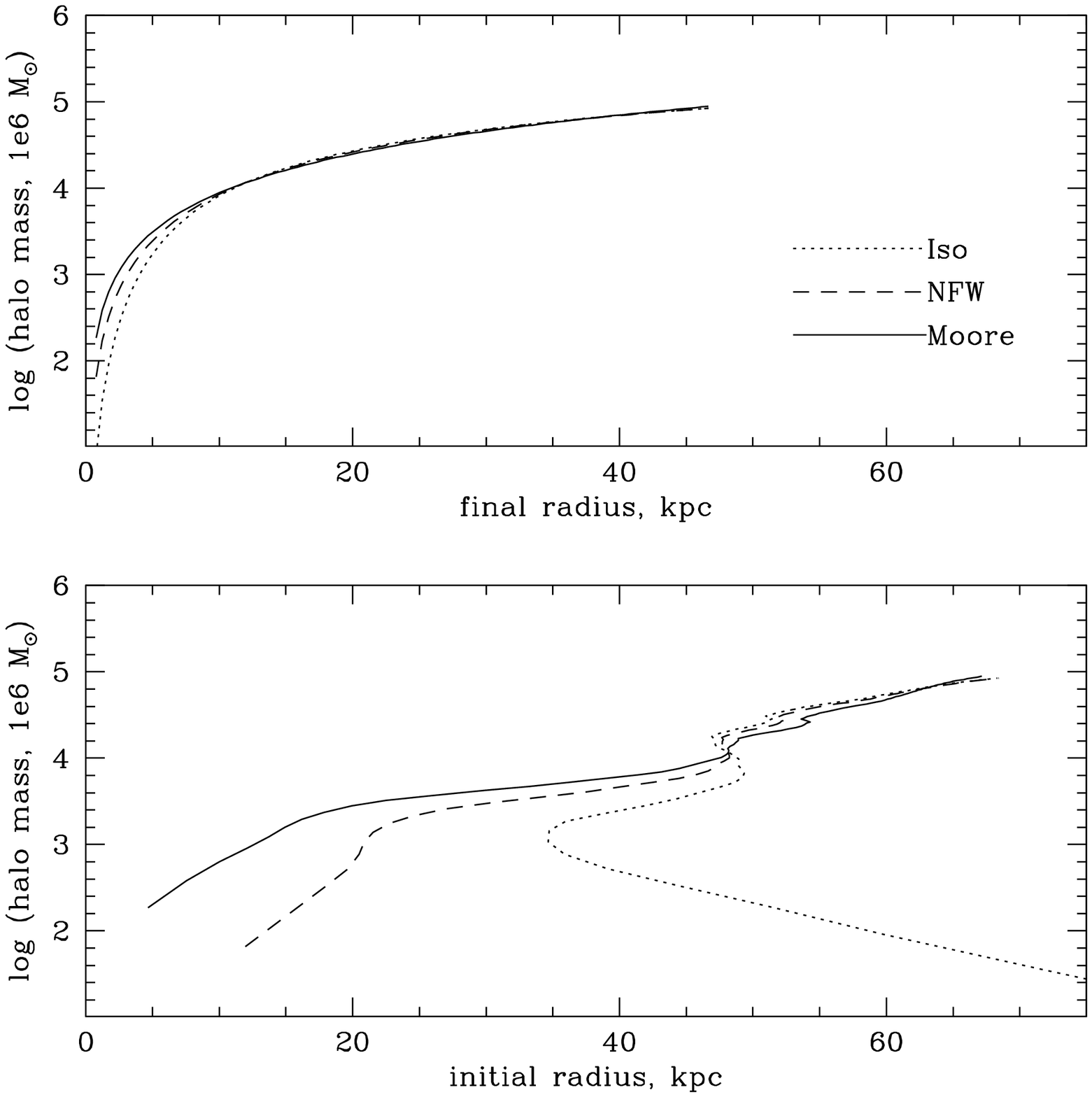}
\end{center}

\caption{Initial and final halo mass profiles in the 
circular-orbit approximation}
\captsize
Upper panel: Final halo enclosed-mass profiles 
${\rm log}~M_{f,halo}(r_f)$ determined by fitting
the rotation curve, for three model halo profiles:
isothermal (dotted), NFW-type (dashed), and Moore-type (solid line).
Lower panel: Initial halo enclosed-mass profiles 
${\rm log}~M_{i,halo}(r_i)$ obtained using 
the circular-orbit uncompression formalism.
Line styles as above.  The isothermal model is
unphysical inside 35 kpc (see text).
\label{fig-uncompmass}

\endfig

\section{Discussion II.  Is NGC 4123 typical?}
\label{sec-typical}

The comparisons between the observed velocity field of the bar in NGC 4123 
and the gas-dynamical models show conclusively that: (1) the stellar disk 
of the galaxy is high-mass, in the sense of being close to maximum disk, 
implying that the disk dominates the rotation curve until well outside the 
optical radius of the galaxy; and (2) the bar rotates quickly.
These conditions are required to match the strength and location
of the shock observed in the bar.

We found that NGC 4123 has a disk $I$-band mass-to-light ratio of 2.0 -- 
2.5 (times $h_{75}$), and is dominated by the stellar disk within the 
optical radius.  The best value for disk \mli\ is 2.25, and
$\mli=2.0$ is a hard lower limit.  
Examining the mass models in Figure 8 of Paper I shows 
that for a disk \ml\ = 2.25, the stellar disk contribution
reaches 87\% of the total rotation velocity at the optical 
radius $R_{25}$ of 11.1 kpc, or 76\% of the centripetal 
acceleration at that radius. 
The disk fraction of the mass inside $R_{25}$ is slightly lower,
since disk flattening boosts its contribution to the centripetal
acceleration: the stellar disk is 72\% of the mass inside
$R_{25}$.  The gas disk is responsible for another 5\% of the 
mass inside $R_{25}$.  The dark halo, when assumed to be spherical, is
23\% of the mass inside a sphere of radius $R_{25}$.
Table \ref{table-massfracs} summarizes the mass fractions
for models with $\mli = 2.0$ to 2.5. In all, the stellar disk
is quite dominant inside $R_{25}$.


\begintab[t]
\begin{center}
\begin{tabular}{llrr}

\tableline\tableline
   &   &  \multicolumn{2}{c}{Fraction of} \\
Disk \mli\ & Component &  $V_{rot}^2$ &  Mass \\
\tableline
All     & Gas disk     &  0.06  &  0.05 \\
2.00    & Stellar disk &  0.66  &  0.63 \\
\nodata & Dark halo    &  0.28  &  0.31 \\
2.25    & Stellar disk &  0.76  &  0.72 \\
\nodata & Dark halo    &  0.18  &  0.23 \\
2.50    & Stellar disk &  0.83  &  0.79 \\
\nodata & Dark halo    &  0.11  &  0.16 \\
\tableline
\end{tabular}

\end{center}
\caption{Disk and halo fractions within $R_{25}$}
\label{table-massfracs}
\tablecomments{
\captsize
Fractions of rotation velocity squared and of mass within a
sphere of radius $R_{25}$ (11.1 kpc).  The mass fractions for 
the dark halo assume that it is spherical.
}
\endtab

We note that an error in the 
distance to NGC 4123 would affect our numerical value of \ml, but would 
not affect our estimate of the the degree of maximality of its disk. A 
larger distance $D$ reduces the disk \ml, with $\ml \propto D^{-1}$, but 
maintains the disk contribution to the rotation curve since $L$ goes up.  

The result that NGC 4123 has a high-mass stellar disk and
a low-concentration dark halo is interesting in itself, since 
it strongly
violates a number of nominally-universal predictions for
the properties of dark halos of galaxies, as described in the
previous section.  However, it is clearly important to
know whether the result applies to disk galaxies
in general.

\subsection{The Tully-Fisher relation}

Perhaps the strongest argument that NGC 4123 is typical is
its place on the Tully-Fisher relation.  As described in
Paper I, NGC 4123 is 0.2 mag more luminous than the mean
$I$-band Tully-Fisher relation for galaxies of type Sbc
and later, which has an intrinsic scatter of
0.3 mag (Giovanelli \etal\ 1997).  One could interpret
its overluminosity as a sign that its \ml\ is 20\% off
the mean, or that it has 20\% more luminous
matter than the mean, or as a 10\% overestimate of its 
distance.  None of these possibilities bring NGC 4123
significantly closer to the theoretical models.
In any case the overluminosity is not statistically
significant.

Scenarios in which most galaxies have a substantial 
amount of dark matter inside the optical radius
and NGC 4123 is an outlier are unappetizing.  A galaxy
at the same place on the TF relation but
with $\sim 50\%$ dark matter inside the optical radius
would have to have a stellar \ml\ different by nearly a
factor of 2.  If stellar \ml\ varied by such a large
amount, the TF relation would have a much larger scatter
than is observed.

\subsection{Barred galaxies}

Is NGC 4123 typical of barred galaxies?
Strongly barred galaxies generally show straight, offset dust lanes like 
those in NGC 4123 (see Athanassoula 1992); many examples can be found in the 
standard atlases (\eg\ Sandage 1961; Sandage \& Bedke 1988).  Spectroscopic 
studies of strongly barred galaxies show large velocity 
jumps associated with these dust lanes, as in Figure 5
of Paper I, NGC 1365 (Lindblad \& J\"orsater 1987, Lindblad \etal\ 1996) 
and NGC 1530 (Regan \etal\ 1997).  The dust lanes indicate the
location of shocks, as discussed above and in Paper I.

The position of the dust lanes favors
fairly fast-rotating bars; in slow bars 
the shocks move too far ahead of the bar, as seen in the simulations 
presented here and the simulations of the Milky Way presented by Weiner \& 
Sellwood (1999). The most secure dynamical determinations of bar pattern 
speeds in other galaxies also indicate that NGC 936 and NGC 4596
have fast-rotating bars, with corotation just outside the end of the bar 
(Merrifield \& Kuijken 1995; Gerssen \etal\ 1999;
using the method of Tremaine \& Weinberg 1984). 

The presence of shocks indicates the disk \ml\ is likely to be high; if 
the dark matter halo dominates the rotation curve, the potential is too 
round to produce strong shocks, as can be seen from our $\ml = 1.0$ 
simulation (Figures \ref{fig-vfields} and \ref{fig-cutml}). 
Furthermore, a high disk \ml\ is also implied by 
fast-rotating bars, since otherwise the bar which forms in a 
heavy halo is quickly slowed by dynamical friction, as
discussed in Section~\ref{sec-maxdisk} (Weinberg 1985, Debattista
\& Sellwood 1998).

\subsection{What about unbarred galaxies?}

NGC 4123 is strongly barred.  Since bars can form by instabilities
in massive disks, but the instability can be deterred by a massive
halo, it has been argued that unbarred galaxies require massive halos
(\eg Ostriker \& Peebles 1973; Efstathiou, Lake \& Negroponte 1982).
Conversely one could argue that barred galaxies are preferentially
disk dominated.  However, it takes a lot of 
halo to deter bar formation: as noted by Kalnajs (1987), central bulges 
are more efficient than extended dark halos at stabilizing disks.  Adding 
more halo mass in the form of spherical shells at large radius has zero 
effect on the dynamics of the matter interior to those shells, so 
stabilizing a disk with a halo requires cranking up the central density.
Additionally, bars can form in disks with massive halos 
(see \eg Combes \& Sanders 1981, Debattista \& Sellwood 1998). 
But is NGC 4123 a self-selected exception by 
virtue of being strongly barred?

First, relations which claim universality should clearly apply to 
unexceptional galaxies, and NGC 4123 is quite ordinary, especially given 
its modest rotation width. Some 30\% of galaxies in major catalogs are 
classified as barred, plus some percentage of intermediate types (Sellwood 
\& Wilkinson 1993), and an even larger bar fraction is found from
infrared imaging (Eskridge \etal\ 2000).
It seems unsatisfying to put forward galaxy 
formation scenarios which exclude such a large percentage of observed 
galaxies.

There are no differences in the overall \hi\ properties of barred and 
unbarred galaxies (Bosma 1996). Palunas (1996) finds that maximum disk 
fits to galaxy rotation curves yield a median $I$-band \ml\ of $(2.4 \pm 
0.9) h_{75}$, very close to our value for NGC 4123.  The distributions of 
maximum disk \ml\ for barred and unbarred galaxies in Palunas's sample are 
indistinguishable (Sellwood 1999).

We are unaware of any evidence that barred and unbarred galaxies fall on 
different Tully-Fisher relations. Many Tully-Fisher surveys tend to avoid 
selecting barred galaxies.  However, Figure 1 of Syer, Mao \& Mo (1999) plots 
a measure of total (disk+halo) mass-to-light ratio versus central surface 
brightness -- a variant of the Tully-Fisher relation -- for the 2446 
galaxies for which Mathewson \& Ford (1996) obtained rotation curves and 
$I$-band photometry.  175 of these galaxies were classified as barred.  
There is no difference in the distribution of barred versus unbarred 
galaxies, implying that the barred and unbarred galaxies do have similar 
luminosities at the same velocity width (see also Sellwood 1999).  
Syer \etal\ nevertheless argued that 
disk galaxies should have low mass disks, $\ml \lesssim 1.4h_{75}$, in 
order to satisfy a stability criterion designed to deter bar formation.

It is extremely difficult to understand how barred and unbarred 
galaxies could lie on the same T-F relation and have similar maximum disk 
\ml\ and \hi\ properties, yet have true disk 
mass-to-light ratios varying by a factor of 2-3.  
In fact, Syer \etal\ themselves do not favor such 
an approach -- they argued that perhaps disks have a 
critical value of \ml, so that bars do not form spontaneously, but can 
form when induced by
\eg\ encounters between galaxies.  This argument is not
particularly compelling, especially in light of the high disk \ml\ of 
NGC 4123 and the dynamical friction argument for high disk \ml\
in most barred galaxies (Debattista \& Sellwood 1998).

Relying on extremely dominant halos to deter bar formation leads  to an 
division between barred and unbarred galaxies for which there is no 
evidence, and in which the existence of intermediate types is hard to 
understand.  Bar formation can be deterred by other 
means than a heavy halo; Kalnajs (1987) emphasized the importance of 
bulges in stabilizing disks
(see also Sellwood \& Evans 2000).  
Bars can be deterred by a central mass dense 
enough to cause an inner Lindblad resonance  (Toomre 1981) or destroyed by 
a relatively small mass accumulation at the center of the bar, which can 
arise naturally as the bar torque drives gas to the center (Pfenniger \& 
Norman 1990, Hasan \etal\ 1993, Friedli \& Benz 1993, Norman \etal\ 1996). 
Many unbarred galaxies with large bulges may have once had bars, and 
whether a high-surface-brightness galaxy is barred or unbarred today may 
depend more on its formation history than on its disk/halo ratio (Sellwood 
\& Moore 1999).

The high disk \ml\ of NGC 4123, with its modest rotation width, suggests 
that the common picture in which lower rotation width galaxies are more 
dark matter dominated may not be correct for high surface brightness 
galaxies. Low surface brightness galaxies appear to be quite dark matter 
dominated (de Blok \& McGaugh 1996, 1997; Swaters \etal\ 2000) 
and surface brightness, rather than rotation width, is likely to be 
the sequence along which dark matter content 
varies.  Detailed models of galaxy formation
predict that specific angular momentum, not mass,
is the parameter which controls the baryon collapse factor,
so that a galaxy with larger angular momentum is 
both more spread out (lower surface brightness) and has less baryonic mass 
within a given number of scale lengths (more dark matter dominated, and 
higher total \ml) (Dalcanton \etal\ 1997).


\section{Conclusions}

Comparing the observed non-circular motions in NGC 4123 to simulations of 
gas flow in model potentials, derived directly from the light 
distribution, shows that the galaxy has a high mass disk and a 
fast-rotating bar.  The best match between simulations and data is for an 
$I$-band disk \ml\ of 2.25 and a bar pattern speed \omegap\ of 20 \kmskpc, 
implying a Lagrange radius \rlag\ of 1.35 times the bar semimajor axis
$a$.  The 
acceptable range of the $I$-band disk \ml\ is 2.0 -- 2.5.  The maximum disk 
value is 2.5 for a disk in an isothermal halo, or 2.25 for a disk 
in a power-law halo similar to the NFW profile, so the galaxy is
80\ -- 100\% of the maximal disk. The acceptable range of 
\omegap\ is from 18 to 24, implying that $\rlaga < 1.5$.

Lighter disks can be excluded because they do not produce shocks as strong 
as those observed.  Slower bars can be excluded because the shock occurs 
in a location which does not match the observations: slower bars produce a 
shock which is farther ahead of the bar, in the direction of galactic 
rotation.

The observed shocks occur at the location of straight offset dust lanes of 
the type commonly seen in strong bars (Athanassoula 1992).  The ubiquity 
of these dust lanes suggests that strong bars generally rotate quickly, 
otherwise the shocks would occur in the wrong place.  Fast-rotating bars 
imply near-maximal disks, since in halo-dominated disks the bar is slowed 
by dynamical friction (Weinberg 1985; Debattista \& Sellwood 1998).  Low 
mass disks also do not produce shocks which extend the length of the bar, 
unless the pattern speed is very slow, in which case the shocks are again 
too far ahead of the bar.  These results suggest that strongly barred 
galaxies have high mass disks and fast-rotating bars.

The near-maximal disk of NGC 4123 does not agree with a number of 
predictions which suggest that galaxies with rotation curve height 
$\simeq 130$ \kms\ should be dominated by dark matter inside the
optical radius (\eg Persic \etal\ 1996; Navarro 1998b) 
nor with predictions that disk galaxies must have low \ml\ 
to satisfy stability criteria (\eg\ Efstathiou \etal\ 1982).  The high 
disk \ml\ can be consistent with a power-law NFW-type halo, but the halo 
must be much less dense than predicted by the relations between 
halo concentration and mass found in CDM and $\Lambda$CDM
simulations (Navarro \etal\ 1996; Bullock \etal\ 1999).  When we 
attempt to derive the initial state of the halo from the adiabatic 
compression and circular-orbit formalism commonly used to model
disk collapse within a dark halo (Faber 1981; Blumenthal \etal\ 1986), 
the inferred initial density profile goes as $r^{0.3}$.  This 
unrealistic halo profile indicates a
failure of the circular-orbit approximation.

Though NGC 4123 is barred, there is no sign that barred galaxies are more 
disk dominated than unbarred galaxies, at least for high surface 
brightness galaxies; their Tully-Fisher relations and
total mass-to-light ratios are similar. 
Requiring heavy halos to stabilize disks is not necessary, since bars can 
also be deterred or destroyed by central bulges.  The ultimate question
of why some galaxies are barred and some unbarred remains
unsettled, but dark halo mass is not the controlling factor.

We have shown that the noncircular motions in a barred galaxy can be used 
to determine the \ml\ of its stellar component.  Together with \hi\ 
observations, these can be used to constrain the radial density profile of 
the galaxy's dark matter halo.  While not every barred galaxy is suitable 
for the modeling process outlined here, observations and modeling of 
more galaxies will eventually lead us to determinations of stellar \ml\ 
and dark halo parameters for galaxies as a function of their rotation 
width, surface brightness, and position on the Hubble sequence.

\acknowledgments

We thank Jacqueline van Gorkom for her contributions to and
support of this project,
Dick van Albada and Lia Athanassoula for providing the 
fluid-dynamical code and for comments on its use,
Povilas Palunas and Julianne Dalcanton for many
helpful conversations and encouragement, and
Jason Prochaska and the
anonymous referee for comments which improved the manuscript.
This research was supported in part by NSF grant AST 96/17088 
and NASA LTSA grant NAG 5-6037 to JAS, and by NSF grant AST 96-19510 to TBW.
BJW has been supported by a Barbara McClintock postdoctoral
fellowship from the Carnegie Institution of Washington.



\end{document}